\documentclass[preprint,12pt]{imsart}

\usepackage{amsmath,amsfonts,bm}
\usepackage{algorithmic}
\usepackage{algorithm}
\usepackage{array}
\usepackage[caption=false,font=normalsize,labelfont=sf,textfont=sf]{subfig}
\usepackage{textcomp}
\usepackage{stfloats}
\usepackage{url}
\usepackage{verbatim}
\usepackage{graphicx}
\usepackage{cite}
\usepackage{hyperref}
\usepackage{natbib}
\usepackage{cite}
\usepackage{amscd,amsmath,amssymb,amsfonts,latexsym,mathrsfs,amsthm,mathtools}
\usepackage[capitalize]{cleveref}
\usepackage{balance}

\newcommand{\V}[1]{\boldsymbol{\mathbf{#1}}}
\newcommand{\Vx}{\V{x}}
\newcommand{\given}{\!\mid\!}

\DeclareMathOperator{\Cov}{cov}

\newcommand{\iid}{\stackrel{\mathrm{i.i.d.}}{\sim}}

\usepackage{xcolor,colortbl}
\definecolor{c100}{RGB}{40,140,40}  
\definecolor{c036}{RGB}{100,180,100}
\definecolor{c013}{RGB}{160,210,160}
\definecolor{c005}{RGB}{200,230,200}
\definecolor{c001}{RGB}{235,245,235}
\begin{document}

\begin{frontmatter}
\title{Sequential Inference for Gaussian Processes: A Signal Processing Perspective\thanksref{t1}}
\runtitle{Sequential Inference for Gaussian Processes}
\thankstext{t1}{\tiny Accepted for publication in \textit{IEEE Signal Processing Magazine}.
This is the authors' accepted manuscript version.
\\© 2026 IEEE. Personal use of this material is permitted.  Permission from IEEE must be obtained for all other uses, in any current or future media, including
reprinting/republishing this material for advertising or promotional purposes, creating new collective works, for resale or redistribution to servers or lists, or reuse of any
copyrighted component of this work in other works by sending a request to pubs-permissions@ieee.org.}

\begin{aug}

\author[A,B]{\fnms{Daniel}~\snm{Waxman}\ead[label=e1]{dan@basis.ai}},
\author[C]{\fnms{Fernando}~\snm{Llorente}\ead[label=e2]{fllorente@bnl.gov}}
\and
\author[B]{\fnms{Petar M.}~\snm{Djuri\'c}\ead[label=e3]{petar.djuric@stonybrook.edu}}

\address[A]{Basis Research Institute\printead[presep={,\ }]{e1}}

\address[B]{Department of Electrical and Computer Engineering,
Stony Brook University\printead[presep={,\ }]{e3}}

\address[C]{Computing and Data Sciences Directorate, Brookhaven National Laboratory\printead[presep={,\ }]{e2}}
\runauthor{D. Waxman et al.}
\end{aug}

\begin{abstract}
The proliferation of capable and efficient machine learning (ML) models marks one of the strongest methodological shifts in signal processing (SP) in its nearly 100-year history. ML models support the development of SP systems that represent complex, nonlinear relationships with high predictive accuracy. Adapting these models often requires sequential inference, which differs both theoretically and methodologically from the usual paradigm of ML, where data are often assumed independent and identically distributed. Gaussian processes (GPs) are a flexible yet principled framework for modeling random functions, and they have become increasingly relevant to SP as statistical and ML methods assume a more prominent role. 
We provide a self-contained, tutorial-style overview of GPs, with a particular focus on recent methodological advances in sequential, incremental, or streaming inference. We introduce these techniques from a signal-processing perspective while bridging them to recent advances in ML. Many of the developments we survey have direct applications to state-space modeling, sequential regression and forecasting, anomaly detection in time series, sequential Bayesian optimization, adaptive and active sensing, and sequential detection and decision-making. By organizing these advances from a signal-processing perspective, we intend to equip practitioners with practical tools and a coherent roadmap for deploying sequential GP models in real-world systems.
\end{abstract}
\tableofcontents

\end{frontmatter}

\maketitle

\section{Introduction}

A fundamental task of statistical signal processing (SP) is the estimation of a signal that is partially observed, embedded in noise, or both. This simply-stated goal is rather difficult in practice and often relies on assumptions we make about the signal; to make these assumptions explicit, the signal is often represented as a \emph{state-space model}, where an observed \emph{state} (or signal) $f_t$ evolves in time in a way that depends only on $f_{t-1}$, emitting noisy observations $y_t$ that depend only on the current state $f_t$. A common set of assumptions is that the state evolves as a linear function over time with a Gaussian random walk and that observations depend linearly on the state and are embedded in Gaussian noise. These assumptions are rather useful, and the estimation of $f_t$ can be optimally done via the Kalman filter.  At the same time, the assumptions are also fairly restrictive.

In this tutorial, we discuss Gaussian processes (GPs), a methodological tool with a number of desirable properties for the signal estimation task. First, GPs naturally cope with \emph{nonlinear} signals and can handle a broad variety of estimation tasks as a result. They accomplish this by appealing to the more abstract property of \emph{correlation} between samples of a signal and making assumptions about the structure of correlation, rather than on a specific parametric form of the model. Second, GPs naturally support  Bayesian inference and provide quantitative uncertainty estimates in signal estimation and reconstruction. This is a critical feature in many SP-based applications, e.g., in healthcare. Finally, GPs operate over a continuum, i.e., they do not assume a discrete-time model of a signal. They, therefore, can be applied to a wide range of problems that may exhibit irregular sampling.

For all their theoretical elegance, GPs are often marred by computational challenges. In particular, in their classical formulation that is often used in machine learning (ML) and statistics, inference of a GP using $N$ samples of a signal requires the inversion of a dense $N \times N$ matrix, which results in $\mathcal{O}(N^3)$ computational complexity. This is particularly problematic in the SP context, where inference of long time series is typical. Furthermore, the classical GP formulation provides no straightforward way to update our beliefs based on the first $N$ samples after observing the $(N+1)$-th sample without incurring at least $\mathcal{O}(N^2)$ operations. These computational limitations have limited the use of traditional GP inference in sequential or resource-constrained SP  and hindered their widespread adoption.

These computational challenges with scaling data are pernicious, and while particularly problematic in SP, they have been extensively studied in neighboring communities as well. Many algorithms have been proposed in the ML literature for scalable GP inference, with a modern trend toward formulations that are explicitly compatible with sequential and
online operations. While these operations are fundamentally related to statistical SP and may be best understood in that setting, these innovations remain underutilized. 

One such approach relies on basis expansions, which approximate a GP prior using a finite set of features, chosen deterministically or randomly, and converge to the true GP prior as the number of features increases. This converts the complicated function-space inference of a GP into a parametric linear model.
Examples include
random Fourier features (RFF) and Hilbert–space GPs, both of which yield state–space structures that can
be updated in real time through standard filtering equations (and, in particular, Kalman filtering, if the observations are embedded in Gaussian noise). Similarly, sparse and variational approximations summarize the influence of past observations through a compact set of inducing variables to allow tractable inference with constant memory and linear complexity.  

The Markovian formulation of a GP 
provides an alternative perspective by expressing 
many common GP priors as linear stochastic differential equations (SDEs). Although more mathematically elaborate, these models, too, allow exact inference through Kalman filtering and smoothing. 
Each of these
approaches replaces the infinite–dimensional representation of the GP with an equivalent parametric model whose parameters evolve sequentially with data.

From an SP viewpoint, these approximations return the GP framework to familiar territory.
They recover the recursive structure of state–space models while retaining the interpretability and
uncertainty quantification of Bayesian inference.  The resulting algorithms extend classical tools such as
recursive least squares and Wiener filtering into probabilistic forms that remain stable, adaptive, and
computationally efficient.  This synthesis also provides new insight into the relationship between kernel methods and linear systems theory and reveals that many stationary kernels correspond to impulse responses of SDEs.  The connection between continuous–time dynamics, spectral
representations, and sequential estimation lies at the core of the modern SP interpretation
of GPs.

The methods discussed in this article show that GPs, when equipped with appropriate
approximations, constitute practical and scalable tools for real–time inference.  We emphasize their
implementation as linear–Gaussian models, their extension to non–Gaussian likelihoods, and their
integration with ensemble and distributed architectures.  The resulting algorithms operate in
a finite-dimensional parameter space while maintaining consistent probabilistic semantics.
This parametric viewpoint also aligns GPs with recent developments in ML, where deep kernel
learning, variational state–space models, and diffusion processes share a common goal of unifying
functional priors with sequential estimation.

The remainder of this article proceeds as follows.  Section~2 introduces the foundation of GP regression and its finite–dimensional interpretation.
Section~3 discusses various existing approaches in scalable GP inference, such as sparse GP regression, and regression via basis expansions. Section~4 presents the sequential version of the basis expansion case. Section~5 presents Markovian GPs and their equivalence to SDEs,
linking GP inference to Kalman filtering and smoothing.  Section~6 examines variational inference, non–conjugate likelihoods, and
extensions to online ensembles and robust inference.  Subsequent sections explore applications to
forecasting, distributed estimation, and adaptive sensing, and conclude with a comparison to modern deep
learning architectures.  
Our objective is to demonstrate that GPs, formulated through parametric and state–space
approximations, provide computationally efficient and reliable tools for sequential SP.

\section{Gaussian Processes} 

In this section, we introduce GPs in their typical (non-sequential) formulation from the \emph{function-space} perspective, where the model is parameterized by a covariance (kernel) function. This includes a motivating example of a joint Gaussian model of signal outputs, which generalizes to GP regression. We discuss several common kernels and how they can impact the properties of the resulting models. We illustrate that exact inference from this perspective is not scalable in online or sequential settings and requires simplifying approximations in both sequential and non-sequential practices. 

\subsection{A Motivating Example: Multivariate Normal Distributions}

We begin by considering a simple, finite-dimensional example that will naturally extend to the infinite-dimensional case of GPs. Specifically, let us consider a one-dimensional, irregularly sampled, continuous-time signal, $y(t)$. We observe samples at $t \in \{0.0, 1.0, 2.5\}$, with the goal of inferring the unobserved value $y(2.0)$. For now, we assume that our observations are noiseless, but we will remove this assumption later. Our first example of GP inference will arise through building a finite-dimensional, multivariate normal distribution over the signal values $\V{y} = [\, y(0) \enskip y(1) \enskip  y(2.0) \enskip  y(2.5) \,]^\top$, i.e.,
\begin{equation} \label{eq:initial_model_mvn}
    \V{y} \sim \mathcal{N}(\V{\mu}, \V{\V\Sigma}).
\end{equation}
This requires specifying the \emph{mean vector}, $\V\mu$, and \emph{covariance matrix}, $\V\Sigma$. For simplicity, we will assume that $\V\mu = \V{0}$. The covariance matrix is more involved, as it tells us how samples are \emph{related} to one another.  One intuitive assumption is that covariances decay with separation in $t$, for example, $\Cov\big(y(t), y(t')\big) = \exp\!\left(-(t - t')^2\right)$. We may thus revise \cref{eq:initial_model_mvn} as
\begin{equation} \label{eq:revised_model_mvn}
\V{y} \sim \mathcal{N}\!\left(\V{0}, 
\begin{bmatrix}
\cellcolor{c100} 1.000 & \cellcolor{c036} 0.368 & \cellcolor{c005} 0.018 & \cellcolor{c001} 0.002 \\
\cellcolor{c036} 0.368 & \cellcolor{c100} 1.000 & \cellcolor{c036} 0.368 & \cellcolor{c013} 0.105 \\
\cellcolor{c005} 0.018 & \cellcolor{c036} 0.368 & \cellcolor{c100} 1.000 & \cellcolor{c036} 0.779 \\
\cellcolor{c001} 0.002 & \cellcolor{c013} 0.105 & \cellcolor{c036} 0.779 & \cellcolor{c100} 1.000
\end{bmatrix}
\right).
\end{equation}

The core property that makes a Gaussian model, like \cref{eq:revised_model_mvn}, fundamental in GP inference is that marginal and conditional distributions remain analytically tractable once the joint distribution is specified. For clarity, we adopt the following notational conventions: we use $*$ to denote quantities associated with the unobserved $y(2.0)$, and $\odot$ to denote quantities corresponding to the observed signal. In particular, the observed signal is
\begin{equation}
\V y_{\odot} = \begin{bmatrix}y(0) & y(1) & y(2.5)\end{bmatrix}^\top,
\end{equation}
which is described by the covariance matrix
\begin{equation}
\V\Sigma_{\odot\odot} = \begin{bmatrix}
\cellcolor{c100}1.000 & \cellcolor{c036}0.368 & \cellcolor{c001}0.002\\
\cellcolor{c036}0.368 & \cellcolor{c100}1.000 & \cellcolor{c013}0.105 \\
\cellcolor{c001}0.002 & \cellcolor{c013}0.105 & \cellcolor{c100}1.000
\end{bmatrix}.
\end{equation}
According to the properties of the multivariate Gaussian distribution, this submatrix is obtained by deleting the rows and columns corresponding to $y(2.0)$ from $\V{y}$. The covariance \emph{between} the observed and unobserved signals, known as the \emph{cross-covariance}, is therefore
\begin{equation}
\V\Sigma_{\odot *}= \begin{bmatrix}
\cellcolor{c005}0.018 & \cellcolor{c036}0.368 & \cellcolor{c036}0.779 \end{bmatrix}^{\!\top}.
\end{equation}

Following standard results of probability theory, we can derive \emph{exactly} the conditional probability of the unobserved value $y(2.0)$ given the observed samples,
\begin{equation}
p\!\big(y(2.0)\,\big|\,y(0), y(1), y(2.5)\big) = \mathcal{N}(\mu_\text{post}, \sigma_\text{post}^2),
\end{equation}
where  $\mu_{\text{post}}$ is the conditional (posterior) mean of $y(2.0)$ and $\sigma_{\text{post}}^{2}$ is its conditional variance.
The mean $\mu_{\text{post}}$ is a weighted average of the observed samples,
{\renewcommand{\arraycolsep}{1pt}
\begin{equation}
\mu_\text{post} = \underbrace{\V\Sigma_{\odot *}^{\top}\,\V\Sigma_{\odot\odot}^{-1}}_{\text{weights}}\,\V y_{\odot} \;\approx\; \begin{bmatrix}\colorbox{c013}{-0.104} & \colorbox{c036}{0.328} & \colorbox{c036}{0.744}\end{bmatrix}
\begin{bmatrix} y(0) & y(1) & y(2.5)\end{bmatrix}^\top.
\end{equation}
}
The posterior variance  $\sigma_\text{post}^2$ reduces the marginal variance $\V\Sigma_{**}$ by the weighted cross-covariances
\begin{equation}
\sigma_\text{post}^2 = \V\Sigma_{**} - \underbrace{\V\Sigma_{\odot *}^{\top}\,\V\Sigma_{\odot\odot}^{-1}}_{\text{weights}}\,\V\Sigma_{\odot *} \;\approx\; 0.3016.
\end{equation}

\begin{figure}
    \centering
    \subfloat[\label{fig:example_ts}]{
        \includegraphics[width=0.85\linewidth]{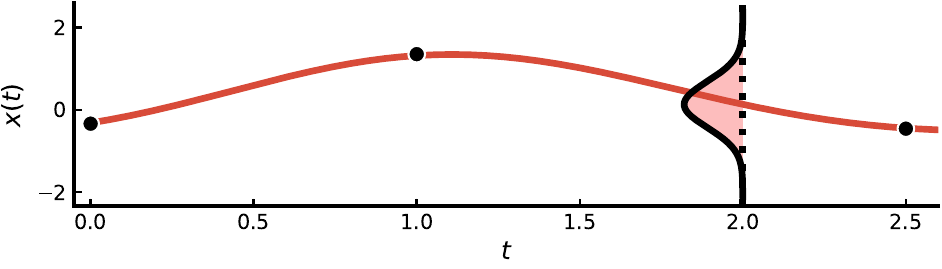}
    }\par\medskip
    \subfloat[\label{fig:example_ts_gp}]{
        \includegraphics[width=0.85\linewidth]{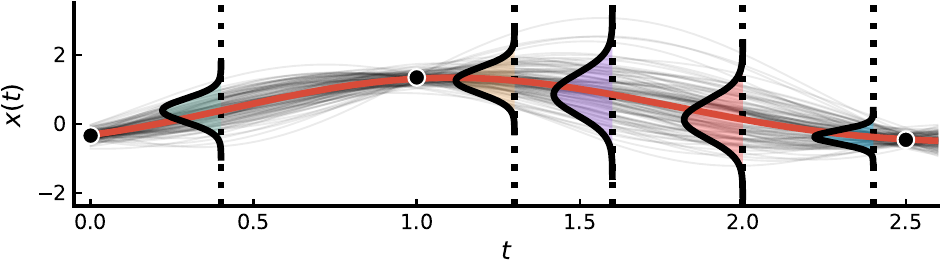}
    }
    \caption{From finite- to infinite-dimensional inference. 
    (\textbf{a}) Inferring the unobserved signal at $t=2$ using the correlations among a finite set of signal values. 
    (\textbf{b}) Extending this idea to the function space: each gray curve represents a random function drawn from the GP posterior that gives rise to the same posterior distribution over $y(2.0)$. GP regression generalizes finite-dimensional Gaussian inference to a continuous domain and provides a unified framework to specify, compute, and sample posterior distributions over functions.}
    \label{fig:example_signal}
\end{figure}

This process is illustrated in \cref{fig:example_ts}. Through our simple probabilistic model, we not only impute $y(2.0)$, but also quantify its uncertainty, which produces a posterior distribution that can be computed via simple linear algebra. In GP regression, the same principle extends to functions: the goal is to infer a posterior probability distribution over {\em functions} that, when covariances are specified appropriately, recovers the same posterior distribution $p(y(2.0) | \V{y}_\odot).$ \emph{GP regression} refers to the Bayesian approach of assigning a GP prior over an unknown function and computing its posterior distribution after observing data. 
This enables both prediction at new inputs and rigorous uncertainty quantification.

\subsection{Extending to Infinite Dimensions: Gaussian Processes}

GP regression is a natural extension of the procedure above for \emph{finite-dimensional random vectors} to the case of \emph{random functions}.  To define a GP, we specify a \emph{mean function} $\mu(t)$ and a \emph{covariance function} $\kappa(t, t')$, which together determine the joint distribution $p(\V{y})$ for any finite collection of observed and unobserved samples $\V{y}$. The defining property of a GP is that this joint distribution, $p(\V{y})$, is {\em always} multivariate Gaussian. More generally, we may consider $d$-dimensional inputs, $\V{x} \in \mathbb{R}^d$, provided that $\mu(\V{x})$ and $\kappa(\V{x}, \V{x}')$ are specified accordingly. Through this construction, we obtain a random function $f \colon \mathbb{R}^d \to \mathbb{R}$, which may be combined with a likelihood $p( y \given f(\V{x}))$ to describe noisy observations. Formally, we write
\begin{align}
    f &\sim \mathcal{GP}(\mu, \kappa), \\ 
    y \given f, \V{x} &\sim p(y \given f(\V{x})).
\end{align}
As the values of $f$ are observed only through $y$, we will sometimes refer to $f$ as the ``latent function.''

A common special case that makes inference analytically tractable arises when the likelihood $p(y \given f(\V{x}))$ is Gaussian, i.e., $y = f(\V{x}) + \varepsilon$ for some $\varepsilon \iid \mathcal{N}(0, \sigma_{\varepsilon}^2)$. In this case, we may proceed using the same conditional normal identities as before to obtain analytical posterior distributions of the latent function  values  at \emph{any collection} of test points, $\V{f}_* = [ f_{*1} \, \cdots, f_{*P} ]$, as
\begin{equation} \label{eq:gp_posterior}
    \V{f}_* \given \V{X}_{\odot}, \V{X}_*, \V{y}_{\odot} \sim \mathcal{N}(\V{\mu}_{\text{post}}, \V\Sigma_\text{post}),
\end{equation}
where the posterior mean and covariance of $\V{f}_*$ are given by
\begin{align}
    \label{eq:gp_posterior_mean}
     \V{\mu}_{\text{post}} 
    &= \V{\Sigma}_{*\odot}\!\left(\V{\Sigma}_{\odot\odot} + \sigma_{\varepsilon}^{2}\V{I}\right)^{-1}\!\V{y}_{\odot}\\
\label{eq:gp_posterior_cov}
    \V{\Sigma}_{\text{post}} 
    &= \V{\Sigma}_{**} 
    - \V{\Sigma}_{*\odot}\!\left(\V{\Sigma}_{\odot\odot} + \sigma_{\varepsilon}^{2}\V{I}\right)^{-1}\!\V{\Sigma}_{\odot *},
\end{align}
where $\V{I}$ is the identity matrix. For the resulting multivariate normal distribution to be mathematically well posed, $\kappa(\V{x}, \V{x}')$ must yield positive semi-definite covariance matrices, and no other mathematical restrictions are imposed. When the likelihood $p(y \given f(\V{x}))$ is non-Gaussian, exact inference is no longer tractable, and we must rely on approximate inference, most commonly variational inference, which we will introduce shortly.

As illustrated in \cref{fig:example_ts_gp}, GP regression generalizes the finite-dimensional Gaussian model of \cref{fig:example_ts} to a posterior distribution over random functions. This means we obtain a \emph{stochastic process} as our posterior, whereby any sample is a function $f \colon \mathbb{R}^d \to \mathbb{R}$.
Whereas \cref{fig:example_ts} depicts inference at a single unobserved point based on correlations among a few samples, \cref{fig:example_ts_gp} shows multiple random functions (gray curves) drawn from the GP posterior. These samples of the posterior function are also known as \emph{realizations} of the GP.
Each realization is consistent with the observed data and reflects the uncertainty specified by the covariance function. 
This visualization illustrates the key conceptual shift: rather than inferring a single value, GP regression infers a \emph{distribution over functions}. Since a GP is just a random function for which any finite collection of points follows a multivariate Gaussian distribution, and since \cref{eq:gp_posterior} is Gaussian, the posterior function $f(\V{x})$ is also a GP.

Of the two functions that must be specified, the covariance function $\kappa(\Vx, \Vx')$ is the  more informative and practically significant. While the mean function $\mu(\V{x})$ provides a time-varying baseline, the covariance function is the component that represents the relationships between pairs of inputs $\Vx$ and $\Vx'$. As a result, and in connection to classical methods in ML, it is often called the \emph{kernel} of the GP. The specific form of $\kappa(\Vx, \Vx')$ represents one of the most influential modeling choices, as it is primarily responsible for determining the shape of the posterior function and even influences the types of approximations or inference methods that can be employed  later. 

There are many potential covariance functions, but several choices stand out as popular and useful. The most common class of kernels relates points according to their Euclidean distance. For example, the squared exponential (SE) kernel (sometimes called the radial basis function (RBF) kernel or exponentiated quadratic (EQ) kernel) assigns an exponentially decaying covariance between points $\V{x}$ and $\V{x}'$,
\begin{equation} \label{eq:se_kernel_def}
    \kappa_\text{SE}(\V{x}, \V{x}') = \sigma_f^2 \exp\!\left(-\frac{\lVert \V{x} - \V{x}' \rVert_2^2}{2\ell^2}\right),
\end{equation}
where $\lVert \mathbf{x} - \mathbf{x}' \rVert_2^2 
= (\mathbf{x} - \mathbf{x}')^\top (\mathbf{x} - \mathbf{x}')$, and $\sigma_f^2$ and $\ell$ are {\em hyperparameters} known as the {\em process variance} and {\em length scale}, respectively. The process variance determines the prior marginal variance of any given function evaluation, while the length scale controls how quickly correlations decay with distance, that is,  how far apart two inputs can be before they lose relevance. 

As noted earlier, the kernel determines many properties of the resulting functions. One important property is their \emph{smoothness}. For example, the SE kernel specifies a prior distribution over infinitely differentiable functions. In many real-world applications, however, this assumption can be overly restrictive,  and the well-known Mat\'ern family of kernels is used instead \citep[pp. 84--85]{rasmussen_and_williams}. Each kernel in this family is parameterized by a \emph{smoothness parameter} $\nu > 0$, which controls the differentiability of the resulting functions: larger values of $\nu$ correspond to smoother functions. The Mat\'ern family is quite general, being defined for any $\nu > 0$, though its form simplifies significantly for half-integer $\nu$. For example, the Mat\'ern-1/2 and Mat\'ern-3/2 kernels correspond to certain non-differentiable and once-differentiable functions, respectively, and are given by
\begin{align}
    \kappa_\text{Mat-1/2}(\V{x}, \V{x}') &= \sigma_f^2 \exp\!\left(-\frac{\lVert \V{x} - \V{x}' \rVert_2 }{\ell}\right); \\ \kappa_\text{Mat-3/2}(\Vx, \Vx') &= \sigma_f^2 \left(1 + \frac{\sqrt{3}\lVert \Vx - \Vx' \rVert_2}{\ell}\right) 
\exp\!\left(-\frac{\sqrt{3}\lVert \Vx - \Vx' \rVert_2}{\ell}\right),
\end{align}
where $\lVert \mathbf{x} - \mathbf{x}' \rVert_2
= \sqrt{(\mathbf{x} - \mathbf{x}')^\top (\mathbf{x} - \mathbf{x}')}$. The Mat\'ern-$\nu$ kernels will prove particularly useful in our later discussion of \emph{Markovian GPs}.

Each of the above kernels is an example of a \emph{stationary} kernel, i.e., a kernel that depends only on the difference $\V{r} = \Vx - \Vx'$. While we will later consider GPs with non-stationary kernels when discussing connections to deep learning, GP priors with stationary kernels remain highly versatile and unlock certain theoretical properties that are useful for approximation, e.g., a spectral decomposition. 

Building on the idea of spectral decompositions, one popular and expressive kernel is the spectral mixtures (SM) kernel \citep{wilson2013gaussian}. The SM kernel represents the \emph{power spectral density (PSD)} of the covariance function, that is, the Fourier transform of a stationary covariance, as a mixture of Gaussians, which allows it to approximate a broad class of stationary covariance functions. We will revisit the SM kernel in subsequent sections. 

\begin{figure}
    \centering
    \includegraphics[width=0.95\linewidth]{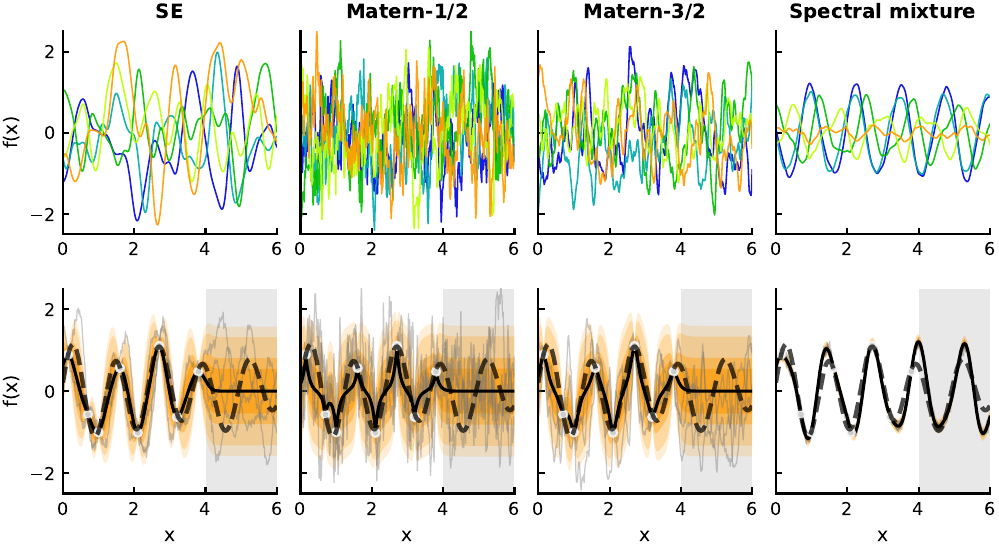}
    \caption{An illustration of prior (top) and posterior (bottom) samples from GPs with different kernels. The posterior plots include an extrapolation region ($x \geq 4$) with no training data. While the SE and Mat\'ern kernels yield smooth or locally correlated behavior that reverts quickly to the prior mean, the spectral mixture kernel captures the underlying quasi-periodic pattern and extrapolates more faithfully beyond the observed region.}
    \label{fig:kernel_examples}
\end{figure}

The choice of kernel is crucial in GP regression, as it affects both the interpolation and extrapolation properties of the resulting model. For example, SE, Mat\'ern-1/2, and Mat\'ern-3/2 kernels each imply different smoothness properties: functions that are realizations from a GP with SE kernel are infinitely differentiable almost surely, whilst realizations from GPs with Mat\'ern-1/2 and Mat\'ern-3/2 kernels are almost surely non-differentiable and once-differentiable functions, respectively. In practical terms, this smoothness determines how rapidly the modeled function can vary between observations. The SE kernels produce very smooth, gently varying functions, while Mat\'ern kernels allow for rougher or more abrupt changes, depending on the value of~$\nu$. More expressive kernels, such as the spectral mixture kernel, can also model quasi-periodic patterns that the SE kernels cannot. We illustrate several different choices of kernels in \cref{fig:kernel_examples}.

Generally speaking, we may combine different covariance functions by either adding or multiplying them \citep[Sec. 4.2.4]{rasmussen_and_williams}. In spatiotemporal settings, we may separate the input as $\V{x} = [ \V{s} \; t ]$, where $\V{s}$ denotes spatial coordinates and $t$ denotes time. If the kernel is expressed as a product $\kappa(\Vx, \Vx') = \kappa(\V{s}, \V{s}') \kappa(t, t')$, it is referred to as \emph{separable} kernel.

GPs are generally sensitive to the specific values of their hyperparameters. The predominant approach for selecting these hyperparameters is \emph{empirical Bayes}, in which the marginal likelihood is maximized. Owing to the simple Gaussian form that the GP reduces to when conditioned on finite data, the resulting log marginal likelihood is simply $\log p(\V{y}_{\odot}) =\log \mathcal{N}(\V{y}_{\odot} \given \V{0}, \V{\Sigma}_{\odot \odot} + \sigma_\varepsilon^2 \V{I}).$
This objective admits closed-form expressions for its gradients and is commonly optimized using non-linear conjugate gradients (as is done in the \texttt{gpml} toolbox for MATLAB \citep{rasmussen_and_williams}), or by gradient descent and quasi-Newton methods (as is more commonly done for software in the Python ecosystem, e.g., \texttt{GPyTorch} \citep{gardner2018gpytorch} or \texttt{GPJax} \citep{pinder2022gpjax}).

\subsection{Troubles in Sequential and Online Estimation}

GPs are now relatively mature tools within the ML and statistics literature and have been successfully applied to many challenging and practically important problems in SP. An important limitation that continues to restrict their broader use, however, is their scalability. While the goal of this article  is to present scalable and sequential GP methods, it is worth examining the source of these challenges.

The primary obstacle to scalability arises from the matrix inversion required in the posterior GP computations, \cref{eq:gp_posterior_mean,eq:gp_posterior_cov}. Although explicit inversion may be avoided by solving equivalent linear systems, computing $\V{\mu}_{\text{post}}$ and $\V{\Sigma}_{\text{post}}$ still incurs an $\mathcal{O}(N^3)$ computational cost and an $\mathcal{O}(N^2)$ memory requirement, where $N$ is the number of data points. 
These quadratic and cubic scalings severely limit the applicability of exact GPs to large data sets. 
There were several early efforts to address these problems (or the related problems in kernel machines) in the SP literature. For example, kernel recursive least squares~\citep{engel2004kernel} includes a recursive form of kernel regression, with a strategy for maintaining a sort of dictionary set of points. The kernel recursive least squares (KRLS) algorithm is interpreted under a Bayesian framework for GP regression by \citet{van2012kernel}, who also provide modifications for time-varying environments. This connection between KRLS algorithms and sequential GP inference is made clear in the tutorial article of \citet{perez2013gaussian}. Unfortunately, apart from windowed approaches, there is typically no way to further reduce the computational cost of exact sequential GP inference beyond $\mathcal{O}(N^2)$ per update in the worst case. 

\section{Approximate Gaussian Process Inference}
As a result of the computational burden of exact GP inference via the kernel formulation, practical sequential inference requires the use of approximate methods. Over the past 15 years, a variety of powerful approximations have been developed that are far more amenable to  online  and sequential settings. In the following, we introduce two such families of approximations that will later be adapted for sequential inference: ({i}) sparse variational methods, which emulate the GP posterior using a set of ``inducing points,'' and ({ii}) basis expansion methods, which approximate the GP through a finite linear model.

\subsection{Sparse Gaussian Process Approximations}

One form of approximation, likely the most common in ML, is a \emph{sparse GP}. Whereas exact GP inference relies on all $N$ training inputs when making predictions, sparse GP approximations seek to summarize the model using a smaller set of $M$ \emph{inducing points}, where $M \ll N$. These inducing points generally consist of \emph{inducing point locations}, $\V{x}_u = [ \V{x}_{u1} \, \cdots \, \Vx_{uM}]$, and a corresponding multivariate distribution over the associated latent function values, 
\begin{equation}
    \V{u} \sim \mathcal{N}(\V{m}, \V{S}). 
\end{equation}

Recall that in exact GP inference, the joint distribution over $\V{y}_\odot$ and  $\V{y}_*$ is used for all predictions. In sparse GP approximations, we assume that the dependence within this joint distribution is \emph{mediated} solely through the inducing variables. This leads to an approximate joint distribution $q(\V{y}_{\odot}, \V{y}_*)$, defined as 
\begin{equation} \label{eq:sparse_gp_approx}
    p(\V{y}_\odot, \V{y}_*) \approx q(\V{y}_\odot, \V{y}_*) = \int_{\V{u}} p(\V{y}_* \given \V{u}) p(\V{y}_\odot | \V{u}) p(\V{u}) \, d\V{u}.
\end{equation}

There are many ways to choose the inducing points and their locations \citep{quinonero2005unifying}, but the most widely used approach today is based on variational inference \citep{titsias2009variational}, which places a distribution $q(\V{u})$ on the inducing points which is later optimized. 

Variational inference (VI) is a general framework for approximate Bayesian inference that consist in approximating the true posterior distribution by a tractable distribution from a simpler family.
The best approximation is found by minimizing the Kullback-Leibler (KL) divergence, which is also intractable. In practice, the minimization of the KL divergence between the approximate and true posterior is equivalent to maximizing the so-called \emph{evidence lower bound} (ELBO) on the marginal likelihood, which leads to a tractable objective for approximate Bayesian inference. Let $p(\V{f}|\V{y})$ and $q(\V{f})$ be respectively the true posterior and its variational approximation, the KL divergence is
\begin{align}
    \text{KL}(q(\V{f})\,\|\,p(\V{f}|\V{y})) &= \int q(\V{f})\log{\frac{q(\V{f})}{p(\V{f}|\V{y})}d\V{f}} \\
    &= \int q(\V{f})\left[\log{q(\V{f})} - \log{p(\V{y}|\V{f})} - 
    \log{p(\V{f})}
    +\log{p(\V{y})}
    \right]d\V{f} \\
    &= \log{p(\V{y})} 
    \underbrace{-\mathbb{E}_{q(\V{f})}\left[
    \log{p(\V{y}|\V{f})}
    \right]
    +
    \text{KL}(q(\V{f}),p(\V{f}))}_{-\mathcal{L}(q)}.
\end{align}
Hence, so long as expectations w.r.t. $q(\V{f})$ are tractable, VI gives us a tractable lower bound on the log-marginal likelihood, i.e., $\log{p(\V{y})}\geq \mathcal{L}(q)$, such that maximizing $\mathcal{L}(q)$ is equivalent to minimizing $\text{KL}(q(\V{f})\,\|\,p(\V{f}|\V{y}))$. Note that $\mathcal{L}(q)$ contains a data-fit term and a regularization term, balancing the quality of the approximation with its complexity. In the context of sparse GPs, the ELBO allows us to optimize the locations of the inducing points and the hyperparameters simultaneously.

In the sparse variational GP framework, we posit a variational distribution over the inducing variables, 
$q(\V{u}) = \mathcal{N}(\V{m}, \V{S})$, 
and optimize the parameters $\V{m}$, $\V{S}$, and the kernel hyperparameters by maximizing the ELBO. 
\citet{titsias2009variational} showed that, when the variational parameters are optimized analytically (the so-called \emph{collapsed} case), the bound simplifies to a closed-form objective that automatically regularizes the inducing points and avoids overfitting. 
\citet{hensman2013gaussian} later generalized this formulation to the \emph{non-collapsed} or \emph{stochastic variational GP} (SVGP), in which the bound is optimized directly with respect to $\V{m}$ and $\V{S}$ and the hyperparameters using stochastic gradient descent and mini-batches of data. 
This stochastic formulation makes GP training scalable to large data sets and facilitates incremental or mini-batch updates.

\subsection{Spectral Gaussian Process Approximations}

Another approach to scalable GP regression, perhaps more reminiscent of traditional SP methods, is based on spectral approximation. Spectral methods aim to emulate the GP by matching its power spectral density (PSD). As previously mentioned in our discussion of the spectral mixture kernel, the PSD exists for GPs with stationary kernels, a result known as Bochner's Theorem \citep[pp.~82]{rasmussen_and_williams}. In this case, we may rewrite the kernel $\kappa(\Vx, \Vx')$ as $\kappa(\V{r})$. Subject to mild measurability constraints, the PSD of the kernel function is given by the Fourier transform of $\kappa(\V{r})$ \citep[pp.~82]{rasmussen_and_williams}.

For many common choices of kernel, the resulting PSD has a simple form and is easy to sample from. For example, the SE kernel results in a Gaussian PSD, up to a scaling factor, and Mat\'ern-1/2 corresponds to a heavier-tailed Cauchy distribution. More expressive kernels, such as those from the MHM \citep{dowling2021hida} or SM families, are carefully constructed to have convenient spectral representations (in both of these cases, via mixture distributions).

The fact that many common kernels induce GP priors with an easy-to-sample PSD motivates a particular parametric approximation of the GP. In particular, the existence of the PSD implies a spectral representation of the kernel, 
\begin{equation} \label{eq:wiener_khinchin_integral}
    \kappa(\V{x}, \V{x}') = \int_{{\mathbb{R}}^D} \exp(i \V{s}^\top (\V{x} - \V{x}') ) S(\V{s}) \, {d}\V{s},
\end{equation}
where $S(\V{s})$ is the PSD of the covariance function, and $i = \sqrt{-1}$ denotes the imaginary unit. Since $S(\V{s})$ is easily sampled from, we create an approximate, parametric prior via direct Monte Carlo integration of \cref{eq:wiener_khinchin_integral}. 

As the PSD of the kernel is given by its Fourier transform, the resulting approximate inference method (via na\"ive Monte Carlo) is known as the \emph{random Fourier feature (RFF)} approximation, introduced for frequentist kernel machines by \citet{rahimi2007random}. Application to GPs, from the perspective of Bayesian trigonometric regression, known as the \emph{sparse spectrum GP}, was introduced by \citet{lazaro2010sparse}. To avoid confusion with state-space GP formulations, as well as a version of the RFF-GP using optimized frequencies, also introduced by \citet{lazaro2010sparse}, we will refer to the resulting approximation as the ``RFF-GP''.

The RFF-GP makes only marginal refinements to a direct Monte Carlo approximation of \cref{eq:wiener_khinchin_integral} by exploiting the symmetric nature of the PSD of a stationary kernel. In the resulting approximation, $F/2$ spectral frequencies $\V{s}_1, \dots, \V{s}_{F/2}$ are sampled from $S(\V{s})$, and a linear basis expansion established,
\begin{equation} \label{eq:rff_gp_defn}
    \V{\varphi}(\V{x}) = \sqrt{\frac{2}{F}} [ \sin(\V{x}^\top \V{s}_1) \, \cos(\V{x}^\top \V{s}_1) \, \cdots \, \sin(\V{x}^\top \V{s}_{F/2}) \, \cos(\V{x}^\top \V{s}_{F/2})]^\top \in \mathbb{R}^F.
\end{equation}
Thus, the GP prior $f \sim \mathcal{GP}(0, \kappa)$ is reduced to a linear model,
\begin{equation} \label{eq:rff_gp_model}
    f(\V{x}) \approx \V{\varphi}(\Vx)^\top \V\theta, \quad \V\theta \stackrel{\mathrm{i.i.d.}}{\sim} \mathcal{N}(0, \sigma_f^2).
\end{equation}
The reduction to a linear model is particularly useful for sequential estimation, as it opens the door to the application of many well-established tools from classical SP.

The use of a direct Monte Carlo approximation of the kernel is conceptually simple but can be inefficient, particularly when $\Vx$ is low-dimensional. For example, orthogonal random features \citep{yu2016orthogonal} or quasi-Monte Carlo \citep{avron2016quasi} can improve inference. An orthogonal approach is to abandon random features in favor of deterministic quadrature of \cref{eq:wiener_khinchin_integral}.

This is the fundamental approach of Hilbert space GPs (HSGPs) \citep{solin2020hilbert}, so-called because of their use of Hilbert space methods from the analysis of partial differential equations (PDEs). As they correspond to certain quadrature rules, HSGPs are simplest to define over 1-dimensional inputs and are extended in a grid-like fashion. Additionally, they must be defined over some compact set for which the approximation is valid.

The resulting approximation of the HSGP is another linear basis expansion \citep{solin2020hilbert}
\begin{equation}
    \phi_k(x) = \left(S\bigg(\frac{k\pi}{2L}\bigg)\right)^{1/2} \frac{\sin\big(\frac{k\pi}{2L}(x + L)\big)}{\sqrt{L}}, \quad k=1, \dots, F,
\end{equation}
where, we recall, $S(s)$ denotes the PSD of the kernel. The HSGP thus inherits many convenient properties for sequential inference, similar to the RFF-GP. Using the grid-like extensions to multivariate inputs, HSGPs suffer from exponentially increasing costs. However, additive approximations (i.e., where $f \sim f_1 + \dots + f_D$, where $f_d$ is a one-dimensional GP with $x_d$ as input) have shown strong empirical performance \citep{solin2020hilbert,waxman2024doebe}.

The HSGP approximation is related to work on the stochastic PDE representation of GPs \citep{lindgren2022spde}, which efficiently solves GP regression problems for a variety of spatial fields. The key difference is that works in the SPDE approach to GPs typically consider a basis expansion around each spatial point $\V{s}$ in a mesh, whilst HSGPs consider a global basis expansion. For more discussion, see Section 7.2 of \citep{lindgren2022spde}.

\section{Sequential Estimation via Basis Expansions} 

Finite basis-expansion approximations of GP priors are particularly powerful for sequential estimation, as they result in tractable finite-dimensional models for which a wide range of recursive estimation methods exist. In the conjugate prior case, these models reduce to linear-Gaussian dynamical systems that are directly amenable to Kalman filtering. In this section, we explore such methods and discuss how the connection to filtering theory extends naturally to non-conjugate likelihoods, dynamic environments, and online ensembling.

\subsection{Exact Inference via Kalman Filtering}

The potential of the RFF-GP as a tool in sequential inference was first noted, to the best of our knowledge, by \citet{gijsberts2013real}, who derived incremental updates to the Cholesky decomposition of the covariance of $\V{\theta}$ in the RFF approximation. From an SP perspective, the existence of such a solution is intuitive. Indeed, the RFF approximation of a GP is simply a Bayesian linear model that naturally allows a trivial state-space representation, 
\begin{align} \label{eq:trivial_state_space_rfgp}
\begin{split}
    \V\theta_{t} &= \V\theta_{t-1} \\ 
    y_t &= \V \varphi(\Vx_t)^\top \V\theta_t + \varepsilon_t,\quad t=1,2,\dots
\end{split}
\end{align}

Together with the prior $\V{\theta}_0 \iid \mathcal{N}(\V{0}, \sigma_f^2 \V{I})$, this state–space formulation yields recursive update rules that are formally identical to those of the Kalman filter. 
In this view, the RFF-based GP behaves as a Bayesian linear dynamical system with a static state vector $\V{\theta}$ and Gaussian observation noise. 
Each new data point incrementally refines the posterior mean and covariance of $\V{\theta}$ and permits efficient online inference without recomputing the full covariance matrix.

\subsection{Adding Dynamics}

In SP applications, it is often necessary to perform online estimation in dynamic environments, where the underlying response function $f(\V{x}_t)$ may itself evolve over time, denoted as $f_t(\V{x}_t)$. 
When using basis expansions of GPs, several strategies can be employed to account for such time-varying behavior. 

The first approach, common in spatiotemporal regression problems, is to incorporate a time-dependent kernel. 
That is, instead of modeling a time-invariant function $f(\V{x})$, we explicitly model a time-varying function $f(\V{x}, t)$~\citep{cressie2011statistics}. 
This formulation has the advantage of preserving the interpretability of the GP prior. 
Moreover, since it retains the static state–space structure of \cref{eq:trivial_state_space_rfgp}, the filtering solution obtained via the Kalman filter is the complete posterior distribution over $\V\theta$; i.e., after a single forward pass over the data, we can predict $f(\V{x}, t_*)$ at arbitrary $t_*$ using the corresponding $p(\V\theta \given y_{1:T})$.

\begin{figure}[t]
    \centering
    \begin{minipage}[t]{0.45\linewidth}
        \centering
        \subfloat[\label{fig:DRFGP_vs_STRFGP_vs_GT}]{
            \includegraphics[width=0.95\linewidth]{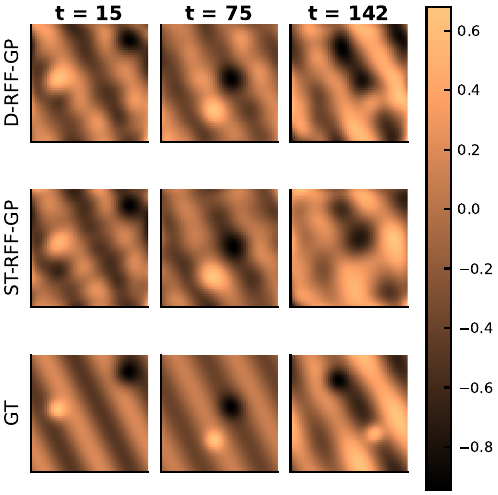}
        }
    \end{minipage}%
    \begin{minipage}[t]{0.45\linewidth}
        \centering
        \subfloat[\label{fig:DRFGP_vs_STRFGP_err}]{
            \includegraphics[width=0.95\linewidth]{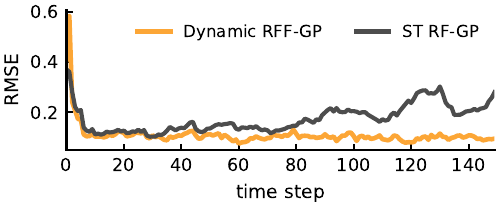}
        }\par\medskip\vspace{-1em}
        \subfloat[\label{fig:DRFGP_vs_STRFGP_f}]{
            \includegraphics[width=0.95\linewidth]{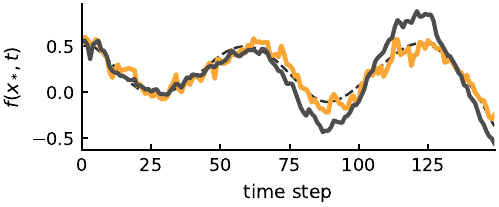}
        }
    \end{minipage}
    
    \caption{Illustrating the relative performance of the dynamic RFF-GP and spatiotemporal RFF-GP. 
    \textbf{(a)} Snapshots of predictions at different points of online estimation of a spatiotemporal field. 
    Note the decline in resolution of the ST-RFGP as time advances. 
    \textbf{(b)} The error over time of each method, measured as RMSE of the mean. 
    \textbf{(c)} Predictions at a particular point in the spatiotemporal field, $x_* = (0.4, 0.6)$.}
    \label{fig:drfgp_vs_strfgp}
\end{figure}

The main drawback of the time-dependent kernel approach to dynamic basis expansions is a gradual loss of spatial resolution as $t$ increases. 
This occurs because the finite set of basis functions must simultaneously represent spatial structure and temporal evolution. 
As time progresses, the fixed $F$-dimensional coefficient vector $\V{\theta}$ must account for an increasing number of effective spatial frames, which causes its representational capacity to become more thinly allocated across time and results in coarser spatial detail. This effect, illustrated in \cref{fig:drfgp_vs_strfgp}, contrasts with the random-walk formulation introduced next, which maintains spatial fidelity by evolving $\V{\theta}_t$ dynamically over time.

An alternative approach is to introduce a nontrivial state–space model for the time-varying coefficients $\V{\theta}_t$. 
The simplest such model assumes that $\V{\theta}_t$ follows a random walk, obtained by adding isotropic Gaussian noise to the parameter vector:
\begin{align} \label{eq:rw_state_space_rfgp}
\begin{split}
    \V{\theta}_{t} &= \V{\theta}_{t-1} + \bm\varepsilon_{\mathrm{rw}, t}, \\ 
    y_t &= \V{\varphi}(\V{x}_t)^\top \V{\theta}_t + \varepsilon_t, \quad t=1,2,\dots
\end{split}
\end{align}
where $\bm\varepsilon_{\mathrm{rw}, t} \iid \mathcal{N}(\V{0}, \sigma_{\mathrm{rw}}^2 \V{I})$. 
This formulation was adopted by \citet{lu2023surrogate} in their so-called \emph{dynamic RFF-GP}, and more generally for basis-expansion models by \citet{waxman2024doebe}. 
In both cases, significant empirical benefits have been observed in time-varying scenarios for both RFF-based GPs and HSGPs.

Under the augmented model~\cref{eq:rw_state_space_rfgp}, we obtain online updates once again via the Kalman filter: 
the predictive step propagates the posterior mean and covariance of $\V{\theta}_t$ through the random-walk dynamics, 
\begin{align}
\hat{\V{\theta}}_{t|t-1} = \hat{\V{\theta}}_{t-1|t-1}, 
\qquad 
\V{P}_{t|t-1} = \V{P}_{t-1|t-1} + \sigma_{\mathrm{rw}}^2 \V{I},
\end{align}
and the measurement update incorporates the new observation $y_t$ through the feature vector $\V{\varphi}(\V{x}_t)$:
\begin{align}
K_t &= \V{P}_{t|t-1}\V{\varphi}(\V{x}_t)\big(\V{\varphi}(\V{x}_t)^\top \V{P}_{t|t-1}\V{\varphi}(\V{x}_t) + \sigma_\varepsilon^2\big)^{-1},\\
\hat{\V{\theta}}_{t|t} &= \hat{\V{\theta}}_{t|t-1} + K_t\!\big(y_t - \V{\varphi}(\V{x}_t)^\top \hat{\V{\theta}}_{t|t-1}\big), 
\qquad 
\V{P}_{t|t} = (\V{I} - K_t\V{\varphi}(\V{x}_t)^\top)\V{P}_{t|t-1}.
\end{align}
These recursive updates efficiently track the time-varying coefficients $\V{\theta}_t$ and, by extension, the evolving function $f_t(\V{x})$, thereby yielding an online Bayesian filtering solution for dynamic GP regression.

\begin{figure}
    \centering
    \includegraphics[width=0.95\linewidth]{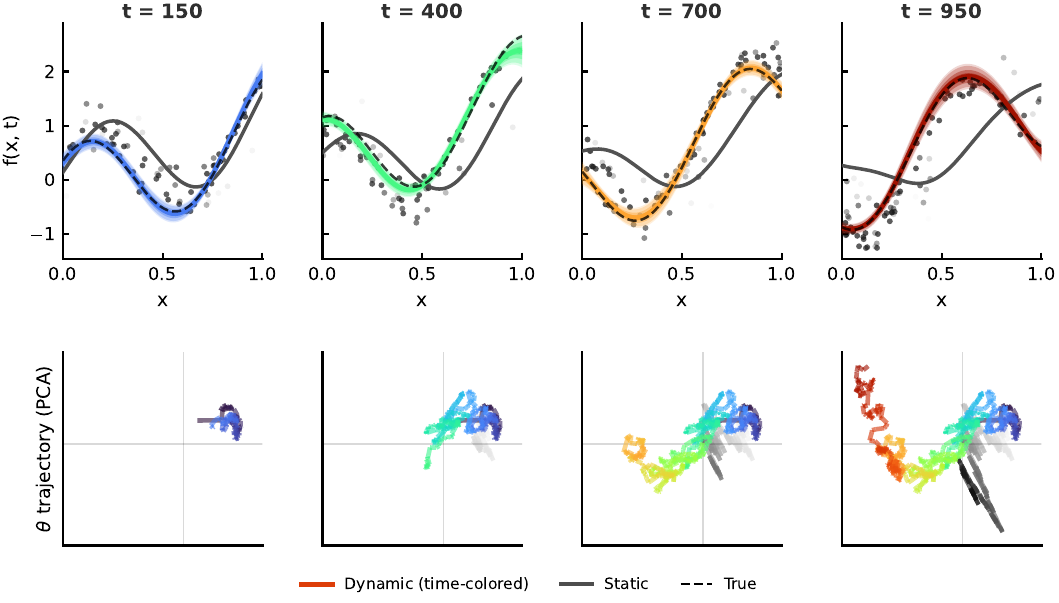}
    \caption{Illustration of the adaptive behavior of an RFF-GP with time-varying parameters. 
{\bf Top}: True time-varying function and corresponding model predictions with (colored) and without (gray) random-walk dynamics. 
Recent data points are shown as darker dots, with older points omitted for clarity. 
{\bf Bottom}: Evolution of the parameter vector $\V{\theta}_t$, visualized through a two-dimensional PCA projection (PC1 vs. PC2) that reveals the gradual adaptation of the model over time. }
    \label{fig:example_drfgp}
\end{figure}

Compared with the spatiotemporal kernel approach, the random-walk state–space model has the advantage of retaining the entire finite-dimensional approximation for a single, time-varying spatial frame. As a result, it tends to produce crisper spatial reconstructions, albeit at the cost of requiring $\mathcal{O}(T)$ storage and an additional backward pass over the data to obtain smoothing estimates. The potential effectiveness of the random-walk formulation of dynamic RF GPs is illustrated in \cref{fig:example_drfgp}.

More generally, the state–space evolution of $\V{\theta}_t$ can be extended to include control inputs and non-isotropic process noise to obtain a variety of time-varying formulations. 
This leads to parameterized dynamics of the form (cf.~\citet{llorente2024dynamic})
\begin{align} \label{eq:complex_state_space_rfgp}
\begin{split}
    \V{\theta}_{t} &= a_t \V{\theta}_{t-1} + \V{u}_t + \varepsilon_{\V{\theta}, t}, \\ 
    y_t &= \V{\varphi}(\V{x}_t)^\top \V{\theta}_t + \varepsilon_t, \quad t=1,2,\dots
\end{split}
\end{align}
where $a_t$ is a scalar autoregressive coefficient, $\V{u}_t$ is a control input, and $\bm\varepsilon_{\V{\theta}, t} \iid \mathcal{N}(\V{0}, \V{C})$ is GP noise with covariance $\V{C}$. 
The random-walk model of \cref{eq:rw_state_space_rfgp} is recovered as the special case of $a_t = 1$, $\V{u}_t = \V{0}$, and $\V{C} = \sigma_{\mathrm{rw}}^2 \V{I}$. 
Alternative parameterizations of these quantities result in a variety of dynamic formulations for time-varying functions discussed in the literature.

One example is \emph{back-to-prior} (B2P) forgetting, first introduced for recursive kernel-based learning~\citep{van2012kernel}. 
B2P forgetting blends the current posterior knowledge with the prior, controlled by a forgetting factor~$\lambda \in [0, 1]$, such that $\lambda = 1$ corresponds to fully trusting the current posterior (no forgetting), and $\lambda = 0$ corresponds to reverting to the prior at each time step. 
B2P forgetting can be recovered from the general state–space formulation in \cref{eq:complex_state_space_rfgp} by setting
\begin{align}
a_t = \sqrt{\lambda}, 
\qquad 
\V{u}_t = \V{0}, 
\qquad 
\V{C} = (1 - \lambda)\sigma_f^2 \V{I}.
\end{align}
This choice introduces controlled diffusion around the prior and allows the model to adapt gradually to nonstationary or time-varying functions while it retains stability and bounded uncertainty.

\subsection{Approximate Inference for Non-Conjugate Likelihoods}

The reduction of the GP model to a linear–Gaussian state–space form has proven extremely convenient for online and sequential inference. 
This perspective is equally useful in the case of non-conjugate likelihoods, such as those arising in GP classification. 
In such cases, we obtain a modified state–space model of the form
\begin{align} \label{eq:nc_state_space_rfgp}
\begin{split}
    \V{\theta}_{t} &= a_t \V{\theta}_{t-1} + \V{u}_t + \varepsilon_{\V{\theta}, t}, \\ 
    f_t &= \V{\varphi}(\V{x}_t)^\top \V{\theta}_t, \\
    y_t &\mid f_t \sim p(y_t \mid f_t), \quad t=1,2,\dots
\end{split}
\end{align}
where, for example, the likelihood $p(y_t \mid f_t)$ may be softmax-based in GP classification or Poisson for count data.

\begin{figure}
    \centering
    \subfloat[\label{fig:drfgp_laplace}]{
        \includegraphics[width=0.23\linewidth]{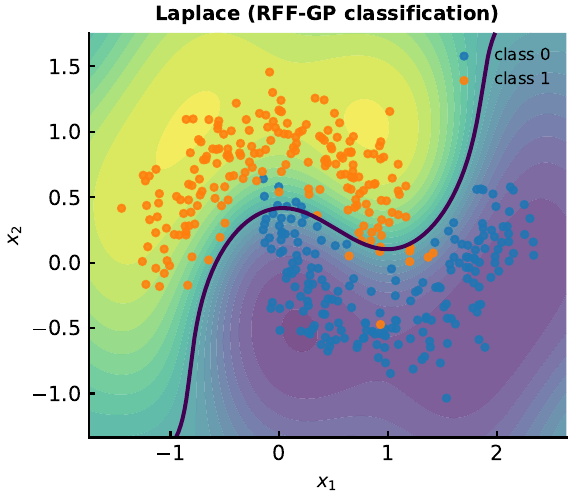}
    }
    \subfloat[\label{fig:drfgp_plf}]{
        \includegraphics[width=0.23\linewidth]{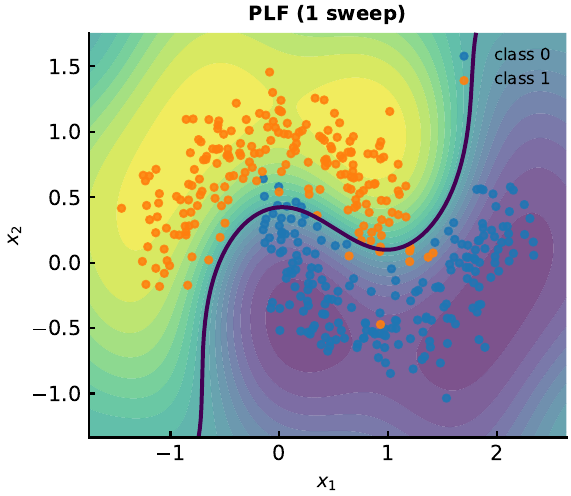}
    }
    \subfloat[\label{fig:drfgp_iplf}]{
        \includegraphics[width=0.23\linewidth]{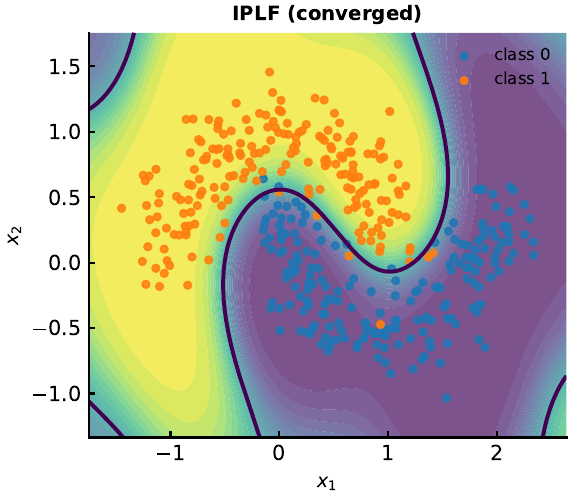}
    }
    \subfloat[\label{fig:drfgp_online_pll}]{
        \includegraphics[width=0.23\linewidth]{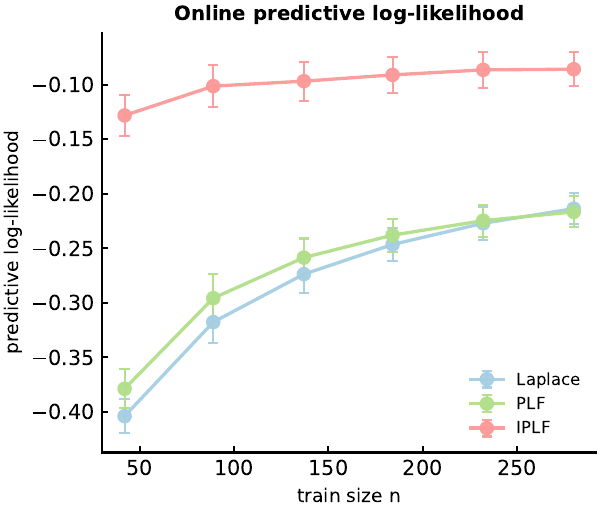}
    }

    \caption{Comparison of inference strategies for non-conjugate likelihoods in an RFF-GP classification task. (a)–(c) Posterior estimates obtained using the Laplace approximation, the posterior linearization filter (PLF), and the iterated posterior linearization filter (IPLF), respectively. (d) Predictive log-likelihoods over time, illustrating the improved accuracy of the IPLF relative to the other two methods, at the cost of higher computational complexity.}
    \label{fig:nonconjugate_comparison}
\end{figure}

For non-conjugate models, a range of filtering-based approximations can be employed. 
These include simple approaches such as the Laplace approximation~\citep{kass1991laplace} and more advanced methods such as posterior linearization~\citep{tronarp2018iterative}. 
A comparison of these techniques is shown in \cref{fig:nonconjugate_comparison}. 
Empirically, iterated posterior linearization filters achieve the best performance but at a substantially higher computational cost than both the standard posterior linearization filter and the Laplace approximation; they may also be more complex to implement.

\subsection{A Bank of Filters Approach}

A persistent challenge in online learning lies in the optimization of hyperparameters. 
GPs, in particular, are highly sensitive to their kernel hyperparameters, and their marginal likelihood is generally non-convex and expensive to optimize. Moreover, efficient online algorithms for hyperparameter adaptation remain limited. As a result, approaches such as the ISSGP of \citet{gijsberts2013real} often rely on pre-trained hyperparameters or costly periodic retraining.

An alternative strategy is to maintain a collection of diverse models in the form of a kernel dictionary and perform online ensembling. This approach transforms the problem of hyperparameter adaptation into a more tractable online model-selection task, for which several effective algorithms exist.

The first work to introduce ensembles of online RFF-GPs was that of \citet{lu2022incremental}, which employed \emph{online Bayesian model averaging} (O-BMA). 
O-BMA is an online and exact variant of the well-known Bayesian model averaging (BMA) framework~\citep{hoeting1999bayesian}, in which models are weighted according to their marginal likelihood. 
Specifically, BMA combines $K$ probabilistic models given data~$\mathcal{D}$, denoted $p_k(y_t \mid \V{x}_t, \mathcal{D})$, for $k = 1, \dots, K$, into a linear mixture,
\begin{equation}
    p_{\mathrm{BMA}}(y_t \mid \V{x}_t) = \sum_{k=1}^K w_{t,k}\, p_k(y_t \mid \V{x}_t, \mathcal{D}),
\end{equation}
where the weights $w_{t,k}$ are proportional to the model evidences,
\begin{equation}
    w_{t,k} = \frac{p_k(\mathcal{D})}{\sum_{j=1}^K p_j(\mathcal{D})}.
\end{equation}
In the online setting, $\mathcal{D}$ corresponds to the available observations $(\V{x}_{1:t-1}, y_{1:t-1})$. 
The weights satisfy a recursive relationship that allows for exact recursive updates,
\begin{equation}
    w_{t+1,k} \propto w_{t,k}\, p_k(y_t \mid \V{x}_{1:t}, y_{1:t-1}).
\end{equation}
This property arises from the recursive factorization of the marginal likelihood,
$p_k(\mathcal{D}_{1:t}) = p_k(\mathcal{D}_{1:t-1})\, p_k(y_t \mid \mathcal{D}_{1:t-1})$,
which allows each model’s weight to be updated incrementally using only its previous value and the likelihood of the latest observation.

From a Kalman filtering perspective, this corresponds directly to the well-known \emph{bank of filters} approach~\citep[Sec.~11.6]{bar2004estimation} for online model combination. 
This connection suggests several potential improvements to the algorithms proposed in the literature, which have typically implemented naïve O-BMA or simple Markov-switching weight updates. 
By contrast, the SP literature has demonstrated the advantages of more flexible switching and weighting mechanisms for combining filter outputs~\citep{el2021particle}.

A more recent alternative to O-BMA is \emph{online Bayesian stacking }(OBS)~\citep{waxman2025bayesian}, which computes the weights~$w_{t,k}$ by solving an online optimization problem. 
Unlike marginal-likelihood maximization, which is non-convex and computationally demanding, OBS seeks the set of weights that maximizes the marginal likelihood of the combined mixture, a convex objective that can be optimized efficiently using established online methods. 
\citet{waxman2025bayesian} demonstrate that OBS achieves favorable performance across several settings, including those involving the online optimization algorithms discussed above.

\subsection{Extensions of Basis Expansions GPs}

The reduction of GPs to linear filters through basis-expansion approximations has proven highly useful for a broad range of SP tasks. In particular, it enables the application of many established techniques from linear filtering theory, extending GPs to new settings. Examples include decentralized inference via consensus algorithms and the information form of the Kalman filter~\citep{llorente2025decentralized}, as well as robust and decentralized inference~\citep{llorente2025robust}, which leverages existing results from robust filtering theory~\citep{chang2017unified}.

The linear formulation of the resulting GP is also advantageous for analytical purposes. 
For instance, accessible bounds on the RFF-GP have led to new theoretical results in Bayesian optimization~\citep{lu2023surrogate} and have been applied in online conformal inference~\citep{xu2025online}.

\section{Markovian GPs} 

In the previous section, we showed that sequential estimation in GPs can be achieved by applying exact Kalman filtering to an approximate linear model. This approach relied on the assumption of a stationary kernel, which permits spectral representations of the GP. Remarkably, for many one-dimensional GPs with stationary kernels, referred to here as \emph{Markovian} GPs, the process can be represented exactly as an SDE, which allows exact inference in linear time through Kalman filter and smoother formulations. 

\subsection{Gaussian Processes as State Space Models}

The central idea of the Markovian GP approach to sequential inference is to represent the GP as a continuous–discrete linear dynamical system. One of the simplest models is the Ornstein–Uhlenbeck (OU) process. This process is described by an SDE over a latent function $f(t)$, which resembles an ordinary differential equation but is driven by a noise process $W(t)$. In the OU case, the deterministic part of the SDE defines a mean-reverting dynamic: the function $f(t)$ is continuously pulled toward zero at a rate $\lambda$, while the stochastic term introduces random fluctuations through the standard Wiener process $W(t)$. The parameter $\lambda$ controls the strength of mean reversion, and $dW(t)$ provides a continuous-time analog of white Gaussian noise with variance scaled to be $q$. Noisy measurements $y_{t_n}$ of the latent function are observed at discrete time points $t_n$. Together, the OU process with discrete observations defines the continuous–discrete state–space model
\begin{align} \label{eq:ou_process}
    \begin{split}
         df(t) &= -\lambda f(t)\,dt + \sqrt{q}\,dW(t), \\ 
         y_{t_n} &= f(t_n) + \varepsilon_{t_n},
    \end{split}
\end{align}
where $q$ and $\lambda$ are hyperparameters, and $\varepsilon_{t_n}$ represents white Gaussian observation noise. It can be shown that this process corresponds exactly to a GP prior with a Mat\'ern–$\tfrac{1}{2}$ kernel when $\lambda = \ell^{-1}$ and $\sigma_f^2 = q / (2\lambda)$, given a prior $f(0) \sim \mathcal{N}(0, \sigma_f^2)$.

Using this representation yields a surprisingly efficient algorithm for GP regression via Kalman filtering. Given data $\mathcal{D}=\{(t_n, y_n)\}_{n=1}^N$, we first discretize the OU state–space model over irregular time steps $\Delta_{t_{n}} \!=\! t_{n}-t_{n-1}$,
\begin{align} \label{eq:ou_process_discretized}
    \begin{split}
        f_{t_{n}} &= e^{-\lambda \Delta t_{n}} f_{t_{n-1}} + \eta_{t_n} \\
        y_{t_n} &= f(t_{n}) + \varepsilon_{t_n},\quad n=1,2,\dots
    \end{split}
\end{align}
where $\eta_{t_n} \sim \mathcal{N}\!\big(0,\, Q_{t_n}\big)$ with 
\begin{align}
Q_{t_n} \;=\; \frac{q}{2\lambda}\,\big(1 - e^{-2\lambda \Delta_{t_n}}\big)
\;=\; \sigma_f^2 \,\big(1 - e^{-2\lambda \Delta_{t_n}}\big),
\end{align}
and $\varepsilon_n \sim \mathcal{N}(0,\sigma_\varepsilon^2)$. 
We then apply Kalman filtering and, if desired, Rauch–Tung–Striebel (RTS) smoothing to obtain \emph{exact} GP inference. For a Markovian GP with state dimension $d$, the total cost is $\mathcal{O}(N d^3)$. Thus, for fixed (small) $d$, the complexity is linear in $N$. 

\begin{figure}
    \centering
    \includegraphics[width=0.85\linewidth]{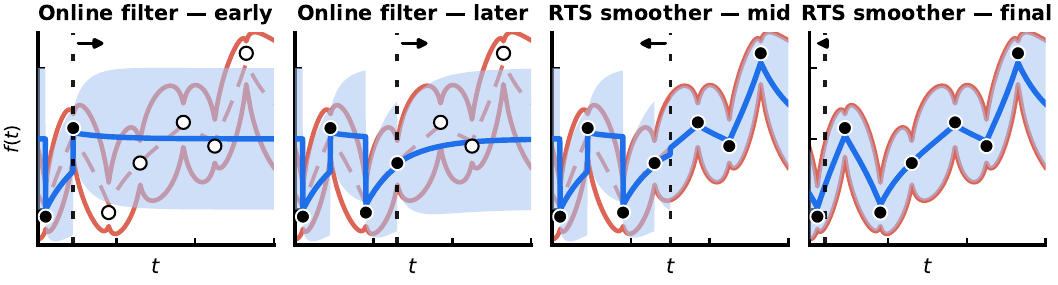}
    \caption{Equivalence of an OU–SDE prior and a GP with a Mat\'ern–$\tfrac{1}{2}$ kernel. 
    We show filtering and smoothing solutions of the OU model alongside the reference GP posterior; illustrated are filtering solutions at an ``early'' observation and a ``later'' observation, and the smoothing solution at a ``mid''-time solution and the ``final'' solution.} 
    \label{fig:example_mat_12_ou_sde}
\end{figure}

This SDE–GP duality naturally raises a question: can other GP priors be represented as linear time-invariant (LTI) SDEs to enable linear-in-$N$ exact inference? 
Although the OU process alone produces rough sample paths, the answer is \emph{yes} by augmenting the state with derivatives.

As a first family, half-integer Mat\'ern kernels admit exact linear-SDE representations. 
For Mat\'ern–$\nu$ with half-integer $\nu$, we define the augmented state to include $f$ and its first $\nu-\tfrac{1}{2}$ derivatives,
\[
\tilde{\V f}(t)=\begin{bmatrix} f(t) & f^{(1)}(t) & \cdots & f^{(\nu-1/2)}(t) \end{bmatrix}^\top.
\]
The driving noise acts only on the highest derivative, with lower orders coupled by a polynomial in $\lambda$. 
(See, e.g., 
\citet{sarkka2019applied}.)
For the Mat\'ern–$\tfrac{3}{2}$ kernel, one convenient continuous–discrete form is \citep[Ex.~12.7]{sarkka2019applied}
\begin{align} \label{eq:mat_32_process}
\begin{split}
    d\tilde{\V f}(t) &= \V F\,\tilde{\V f}(t)\,dt \;+\; \V L\,\sqrt{q}\,dW(t),\\
    y_{t_n} &= \V H\,\tilde{\V f}(t_n) + \varepsilon_{t_n},
\end{split}
\end{align}
with
\[
\V F=\begin{bmatrix} 0 & 1 \\ -\lambda^2 & -2\lambda \end{bmatrix}, 
\qquad 
\V L=\begin{bmatrix} 0 \\ 1 \end{bmatrix}, 
\qquad 
\V H=\begin{bmatrix} 1 & 0 \end{bmatrix},
\qquad 
q \;=\; 4 \lambda^{3}\,\sigma_f^2,
\]
and $\varepsilon_{t_n}\sim \mathcal{N}(0,\sigma_\varepsilon^2)$. Kalman filtering/smoothing
on this two-dimensional state recovers the exact GP posterior for Mat\'ern–$\tfrac{3}{2}$.

A natural question is: \emph{how expressive are the GPs that can be represented in this way?} 
Perhaps surprisingly, they are arbitrarily expressive within the class of stationary GPs. 
Recent work~\citep{loper2021general,dowling2021hida} shows that a GP with any stationary kernel can be approximated arbitrarily well by a simple modification of the Mat\'ern kernel. 
Following~\citet{dowling2021hida}, we refer to these as \emph{mixture Hida–Mat\'ern} (MHM) kernels, although they are also known as \emph{latent exponentially generated} processes in~\citet{loper2021general}.

To construct the MHM family, we begin with the \emph{Hida–Mat\'ern} (HM) kernel, which combines a phase shift~$b$ with a Mat\'ern–$\nu$ kernel:
\begin{equation}
    \kappa_{\mathrm{HM}(b, \nu)}(t, t') 
    = \cos\!\big(b\,|t - t'|\big)\, \kappa_{\mathrm{Mat}\text{-}\nu}(t, t'),
\end{equation}
where all hyperparameters are left implicit. 
We refer to~$b$ as a \emph{phase shift} because it corresponds to a shift by~$b$ in the PSD of the kernel relative to the underlying Mat\'ern kernel. The MHM family is defined as linear mixtures of HM kernels, which provide a Mat\'ern-based analog of the spectral mixture kernel~\citep{wilson2013gaussian}, with the intuitive goal to perform density estimation of the PSD. The resulting state-space formulation is quite similar to the Mat\'ern state space above.

By the linearity of GPs, the sum of kernels corresponds to the sum of independent GPs. 
Thus, for a simple two-component MHM kernel,
\begin{equation}
    \kappa(t, t') 
    = w_1 \cos(b_1 |t - t'|)\,\kappa_{\mathrm{Mat}\text{-}\nu_1}(t, t') 
    + w_2 \cos(b_2 |t - t'|)\,\kappa_{\mathrm{Mat}\text{-}\nu_2}(t, t'),
\end{equation}
the corresponding state–space representation is
\begin{align} \label{eq:sum_markovian_gp}
\begin{split}
    d\tilde{\V{f}}(t) &= 
    \begin{bmatrix} \V{F}_1 & 0 \\[2pt] 0 & \V{F}_2 \end{bmatrix}
    \begin{bmatrix} \tilde{\V{f}}_1(t) \\[2pt] \tilde{\V{f}}_2(t) \end{bmatrix}\!dt
    + \begin{bmatrix} \sqrt{q_1}\V{L}_1 \\[2pt] \sqrt{q_2}\V{L}_2 \end{bmatrix}\!d\V{W}(t), \\[3pt]
    y_{t_n} &= 
    \begin{bmatrix} \sqrt{w_1} \V{H}_1 & \sqrt{w_2} \V{H}_2 \end{bmatrix}
    \tilde{\V{f}}(t_n) + \varepsilon_{t_n}.
\end{split}
\end{align}
Further generalizations follow in the same way and yield block-diagonal covariance matrices that can be computed efficiently and exploited for more efficient filtering and smoothing.

\subsection{Spatiotemporal Regression via Markovian Gaussian Processes}

Although Markovian GPs were originally formulated for temporal processes, they can be naturally extended to spatiotemporal settings. The mathematical strategy remains similar, exploiting the spectral representation of the spatiotemporal kernel to derive an appropriate stochastic partial differential equation (SPDE)~\citep{sarkka2012infinite}. This approach shares deep connections with more general work on the SPDE formalism of GPs \citep{lindgren2022spde}, with the infinite-dimensional filtering presented by \citet{sarkka2012infinite} being an efficient implementation for simple spatial fields (cf. Section 4.3 of \citet{lindgren2022spde}). This general formulation, however, is mathematically involved and typically requires advanced approximate inference methods for SPDEs.

In this tutorial, we focus instead on a simplified and practically useful case that involves \emph{separable} kernels. 
Recall that a spatiotemporal kernel $\kappa\!\left((\Vx, t), (\Vx', t')\right)$ is separable if it can be expressed as the product of a spatial kernel and a temporal kernel, $\kappa_s(\Vx, \Vx')\,\kappa_t(t, t')$. This assumption is convenient because it preserves the Markov property: spatial correlations act only through a filtered and augmented spatial state $\tilde{\V f}_s$. 
The remaining task is to define $\tilde{\V f}_s$ and specify how spatial dependencies are incorporated into the overall model.

In the general case, $\tilde{\V f}_s$ is a functional, that is, an infinite-dimensional spatial state evolving over time, on which infinite-dimensional Kalman filtering and smoothing can, in principle, be applied~\citep{sarkka2012infinite}. 
While this formulation is theoretically elegant and occasionally practical, we again adopt a simpler approach. 

Our final simplification for spatiotemporal GPs is to discretize the spatial domain into a finite set of locations $\Vx_1, \dots, \Vx_{N_s}$. 
The resulting state is then defined by $N_s$ copies of the temporal kernel state $\tilde{\V f}$, i.e.,
\begin{equation}
    \tilde{\V f}_S 
    = \begin{bmatrix} 
        \tilde{\V f}_1^\top & \cdots & \tilde{\V f}_{N_s}^\top 
      \end{bmatrix}^\top.
\end{equation}
The associated SDE has dynamic and feedback matrices formed as the block-diagonal concatenation of $\V F$ and $\V L$. 
Spatial dependencies enter through the diffusion term, whose covariance takes the Kronecker-structured form 
$\V Q_t \otimes \V \Sigma_{SS}$, 
where $\otimes$ denotes the Kronecker product, and $\V \Sigma_{SS}$ is the spatial covariance matrix induced by the spatial kernel.

Assuming that the temporal kernel induces a $d_t$-dimensional state space (using $d_t$ to differentiate between dimensions induced by the temporal kernel and spatial kernel)
, the resulting spatiotemporal model has a $(d_t \times N_s)$-dimensional joint state. Inference through the Kalman filter and smoother then requires $\mathcal{O}(d_t^3 N_s^3 T)$ computations, which yield quadratic savings in~$T$ compared to standard GP inference. In the following, we discuss several extensions that further reduce the cubic scaling with respect to the number of spatial locations~$N_s$.

\subsection{Extensions of Markovian Gaussian Processes}

As with basis-expansion approximations, much of the recent progress in extending Markovian GPs arises from their convenient formulation as linear–Gaussian systems, which allows the direct use of tools from SP. 

A primary challenge in this setting involves non-Gaussian likelihoods. 
Analogous to the basis-expansion case (cf.~\cref{fig:nonconjugate_comparison}), standard nonlinear filtering and smoothing methods can be applied, such as posterior linearization~\citep{garcia2019gaussian}. 
The widespread use of GP classification has even inspired the development of new approximate inference methods for state-space models, such as state-space expectation propagation~\citep{wilkinson2020state}. Variational inference methods are also effective and represent the state of the art in large-scale spatiotemporal applications~\citep{hamelijnck2021spatio} and in modeling count data, including neuroscience applications~\citep{dowling2023linear}. 

Another major challenge for Markovian GPs, especially in spatiotemporal settings, is scalability. Although Markovian formulations avoid the cubic scaling in time typical of standard GPs, they still exhibit $\mathcal{O}(N_s^3)$ complexity in the number of spatial locations, which can become prohibitive. One approach to avoid this is the spatiotemporal sparse variational GP, which introduces a grid of $M_s$ fixed spatial inducing points and achieves $\mathcal{O}(M_s^2 N_s T)$ training and inference costs~\citep{hamelijnck2021spatio}. 
While such models are not sequential in the strict sense of being recursive or online, they still perform inference sequentially in time and thus retain favorable scaling properties. 
Markovian GPs have also motivated new computation-aware filtering and smoothing algorithms~\citep{pfortner2025computation}, which leverage probabilistic linear algebra for scalable inference. 

Finally, efforts to make Markovian GPs robust have benefited directly from the robust filtering literature. 
Simple approaches include rejection filtering~\citep{bock2022online,waxman2024lintel}, while more recent methods employ generalized Bayesian inference to achieve efficient and outlier-robust spatiotemporal GPs~\citep{laplante2025robust}. We will further discuss robustness, and its implications for SP, in the following section.

\section{Sequential Estimation in Sparse (Variational) Approximations} 

Exact GP inference remains computationally expensive for large datasets. 
Sparse and variational approximations overcome this limitation by representing the latent function 
through a compact set of inducing variables that summarize the information from past data. 
In sequential settings, these low-dimensional representations can be updated recursively to allow for efficient online inference without revisiting earlier observations.

When data arrive sequentially, these sparse representations can be updated online rather than 
recomputed from scratch.  This section introduces the main ideas of sequential inference 
within sparse and sparse-variational GPs.  We show how the algebra of 
Gaussian conditioning allows recursive updates of the inducing-point statistics, 
discuss connections to classical filtering concepts, and describe modern variational 
formulations that preserve scalability in both time and data size.

\subsection{Sequential Estimation in Sparse Approximations}

The classical \emph{sparse GP} approximation assumes that the latent function values
\begin{align}
\V f = [f(\Vx_1), f(\Vx_2), \dots, f(\Vx_N)]^\top
\end{align}
depend on a much smaller set of inducing variables
\begin{align}
\V u = [f(\Vx_{u1}), f(\Vx_{u2}), \dots, f(\Vx_{uM})]^\top, 
\qquad M \ll N,
\end{align}
located at inducing inputs $\Vx_u = \{\Vx_{u1}, \dots, \Vx_{uM}\}$.  
The joint prior is
\begin{align}
p(\V f, \V u) = 
\mathcal{N}\!\left(
\begin{bmatrix} 
\V 0 \\ \V 0 
\end{bmatrix},
\begin{bmatrix}
\bm{\Sigma}_{ff} & \bm{\Sigma}_{f u} \\
\bm{\Sigma}_{u f} & \bm{\Sigma}_{u u}
\end{bmatrix}
\right),
\end{align}
where $\bm{\Sigma}_{u u} = [\kappa(\Vx_{u i}, \Vx_{u j})]_{i,j}$ 
is the covariance among inducing inputs, 
$\bm{\Sigma}_{f u} = [\kappa(\Vx_i, \Vx_{u j})]_{i,j}$ 
is the cross-covariance between data and inducing inputs, 
and $\bm{\Sigma}_{ff}$ is the full covariance of the observations.  
Conditioning on $\V u$ yields
\begin{align}
p(\V f \mid \V u) 
= \mathcal{N}(\bm{\Sigma}_{f u}\bm{\Sigma}_{u u}^{-1}\V u, \, \bm{Q}_{ff}),
\end{align}
where $\bm{Q}_{ff} = \bm{\Sigma}_{ff} - \bm{\Sigma}_{f u}\bm{\Sigma}_{u u}^{-1}\bm{\Sigma}_{u f}$ 
represents the residual covariance.  All dependence on the data enters only 
through $\V u$, which becomes the effective state of the approximation. This greatly reduces the computational burden of prediction, as we require the inverse of an $M \times M$ matrix, rather than an $N \times N$ matrix, where $M \ll N$. 

When data arrive sequentially, the posterior over $\V u$ can be updated recursively.  
Suppose at time $t-1$ we have
\begin{align}
p_{t-1}(\V u) = \mathcal{N}(\V m_{t-1}, \V S_{t-1}).
\end{align}
The new observation $(\Vx_t, y_t)$ introduces a likelihood
\begin{align}
p(y_t \mid f_t) = \mathcal{N}(y_t; f_t, \sigma_\varepsilon^2),
\end{align}
and the marginal $f_t$ is linearly related to $\V u$ through 
$f_t = \V k_t^\top \bm{\Sigma}_{u u}^{-1}\V u + \varepsilon_t$,  
where $\V k_t = [\kappa(\Vx_t, \Vx_{u1}), \dots,$ $ \kappa(\Vx_t, \Vx_{uM})]^\top$.  
Using Gaussian conditioning identities, the posterior mean and covariance of $\V u$ follow
\begin{align}
\V K_t &= \V S_{t-1}\V h_t(\V h_t^\top \V S_{t-1}\V h_t + \sigma_\varepsilon^2)^{-1},\\
\V m_t &= \V m_{t-1} + \V K_t (y_t - \V h_t^\top \V m_{t-1}),\\
\V S_t &= \V S_{t-1} - \V K_t \V h_t^\top \V S_{t-1},
\end{align}
where $\V h_t = \bm{\Sigma}_{u u}^{-1}\V k_t$.  
These updates are mathematically identical to those of a Kalman filter with state $\V u$.  Each step requires $\mathcal{O}(M^2)$ computation, which enables online inference without the need to revisit previous observations.

This correspondence with the Kalman filter provides an intuitive interpretation: 
the inducing-point vector $\V u$ serves as the hidden state, 
and each new observation performs a measurement update on that state.  
The matrices $\bm{\Sigma}_{u u}$ and $\V k_t$ act as prior and observation covariances, respectively.  
Under this viewpoint, the sparse GP operates as a linear--Gaussian dynamical system 
with a trivial transition model $\V u_t = \V u_{t-1}$. 

This perspective highlights both the strengths and limitations of the approach.  
Updates are exact within the approximation, but the inducing inputs locations $\Vx_u$ remain fixed. They do not adapt to new data unless explicitly optimized.  
Early placement of $\Vx_u$ therefore heavily influences long-term performance.  
Adaptive selection or relocation of inducing inputs can improve accuracy, 
though at the cost of additional optimization.

Sparse approximations differ from subset-of-data (SoD) or subset-of-regressor (SoR) approaches.  
In subset methods, only a portion of past observations contributes to future predictions, 
whereas in sparse GPs the inducing inputs interpolate the full posterior.  
Sequential updates thus preserve the information content of all past data rather than 
discarding it, which makes sparse GPs well suited for long-term estimation tasks 
in slowly varying systems.

\subsection{Sequential Estimation in Sparse Variational Approximations}

The probabilistic formulation above assumes exact Gaussian identities.  
In practice, the marginal likelihood $p(\V y)$ is rarely tractable, 
especially for non-Gaussian likelihoods $p(y_t \mid f_t)$.  
Variational inference introduces a principled way to approximate this posterior 
while maintaining scalability.

In the \emph{variational sparse} GP (VSGP), we approximate the true posterior 
$p(\V u \mid \V y)$ by a Gaussian variational distribution
\begin{align}
q(\V u) = \mathcal{N}(\V m, \V S).
\end{align}
The evidence lower bound (ELBO) is
\begin{align}
\log p(\V y) \ge 
\sum_{t=1}^T \mathbb{E}_{q(f_t)}[\log p(y_t \mid f_t)]
- \mathrm{KL}\big(q(\V u)\,\|\,p(\V u)\big),
\end{align}
where $q(f_t)$ denotes the marginal of $f_t$ under $q(\V u)$.  
Optimizing this bound yields both the posterior parameters $(\V m, \V S)$ and the kernel hyperparameters.

Sequential inference follows by decomposing the ELBO over time.  When a new observation $(\Vx_t, y_t)$ arrives, we add a new term to the ELBO and update $(\V m, \V S)$ using Kalman filter-like recursion \citep{schurch2020recursive}, i.e., 
\begin{align}
\V S_t^{-1} &= \V S_{t-1}^{-1} + \V H_t^\top \V R_t^{-1}\V H_t,\\
\V S_t^{-1}\V m_t &= \V S_{t-1}^{-1}\V m_{t-1} + \V H_t^\top \V R_t^{-1} \V y_t,
\end{align}
where $\V H_t$ and $\V R_t$ are determined by the local likelihood model. These relations resemble recursive least squares but arise from the variational objective and allow online updates of $(\V m, \V S)$ without revisiting past data.

\begin{figure}
    \centering
    \includegraphics[width=0.95\linewidth]{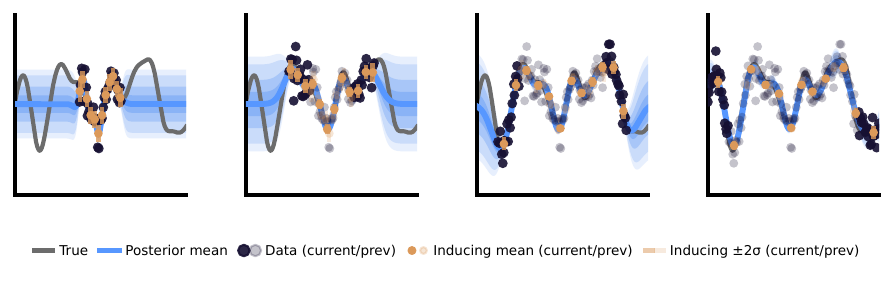}
    \caption{An illustration of the streaming VSGP (via the implementation provided by \citet{chang2023memory}): as new data arrive (black) the GP approximation adapts online, distributing the inducing locations (light and darker yellow) along the interval in order to produce an accurate approximation.} 
    \label{fig:example_bui}
\end{figure}

In the \emph{streaming VSGP} \citep{bui2017streaming}, the current posterior $q_{t-1}(\V u)$ acts as the prior for the next step, and the new posterior $q_t(\V u)$ minimizes
\begin{align}\label{pre-Bui}
\mathrm{KL}\!\left(q_t(\V u)\,\|\,p(\V u \mid \mathcal{D}_{1:t})\right)
\approx
\mathrm{KL}\!\left(q_t(\V u)\,\bigg\|\,\frac{p(y_t \mid \V u)\,q_{t-1}(\V u)}{Z_t}\right),
\end{align}
where $Z_t$ is a normalization constant.  
This formulation enables efficient one-pass updates of the variational parameters.  In the simpler scenario, the inducing inputs $\Vx_u$ remain fixed (along with the hyperparameters) and maintain constant memory even as the data stream grows. 
\citet{bui2017streaming} formalize the intuition in Eq. \eqref{pre-Bui} but consider the more challenging problem of online updating for the inducing inputs and hyperparameters. Denoting with $\theta$ the set of hyperparameters, they consider the {\it intermediate} target density to be instead $\widehat{p}(f|\mathcal{D}_{t-1},y_t, \theta_{t-1},\theta_t)
\propto
p(f|\theta_t)\frac{p(y_t \mid f)\,q_{t-1}(f)}{p(f|\theta_{t-1})}$, hence effectively accounting for the new $\theta_t$ in the $q_t(f)$, with possibly different inducing locations ${\bf x}_{u,t}\neq {\bf x}_{u,t-1}$.
An instance of the streaming VSGP by \citet{bui2017streaming} is shown in Figure 
\ref{fig:example_bui}.
\citet{chang2023memory} propose a memory-based dual-SVGP that augments the inducing-point posterior summary with a small, actively selected buffer of past data, helping sequential conditioning and hyperparameter/inducing-input adaptation track the offline ELBO more closely.

The streaming VSGP also connects to natural-gradient descent in the space of variational parameters \citep{khan2017conjugate,hamelijnck2021spatio}.  Because $q(\V u)$ is Gaussian, its natural parameters are $(\V S^{-1}\V m, \V S^{-1})$, and small updates along these directions correspond to moment matching.  This relationship links online variational inference to recursive information filters in classical SP \citep{opper2009variational}.

\paragraph*{Non-Gaussian likelihoods and Robustness}
The variational formulation easily extends to non-Gaussian likelihoods. The expectation $\mathbb{E}_{q(f_t)}[\log p(y_t \mid f_t)]$ 
occasionally has strong approximations available (e.g., in the case of logit or probit models), or numerically otherwise.  
Sequential updates depend only on the local gradients of this expectation.  
Robust likelihoods, such as Student-$t$ or Huber forms, reduce sensitivity 
to outliers without altering the recursive structure.

A further advantage of the variational approach is its modularity.  
It can combine with structured likelihoods, multi-output GPs, or hierarchical priors 
without changing the recursive pattern.  
This flexibility has made online variational GPs the preferred method 
in complex spatiotemporal models that demand both scalability and adaptability.

The per-step complexity of sequential variational updates is $\mathcal{O}(M^3)$ 
due to matrix inversions in $\bm{\Sigma}_{u u}$ or $\V S$. Because $M$ is fixed and typically small, the total cost grows linearly with time.  
Mini-batches of size $B$ reduce this to $\mathcal{O}(B M^2)$ per update.  
Only the parameters $(\V m, \V S)$ and the kernel matrices associated with $\Vx_u$ 
are stored, so memory remains constant, unlike the $\mathcal{O}(N^2)$ 
storage requirement of exact GPs.

The algebraic similarity between online variational GPs and Kalman filtering is not accidental.  Both maintain a Gaussian belief that is updated through the integration of new information.  The difference lies in interpretation: Kalman filtering assumes a known generative model, whereas variational inference optimizes a lower bound that approximates it.  The update of $(\V m, \V S)$ parallels the correction step of an information filter, with $\V S^{-1}$ acting as the information matrix.

This analogy helps build intuition: we may view the variational updates as approximate filtering, where the innovation term is replaced by a variational expectation.  This viewpoint unifies probabilistic inference and adaptive SP within a common mathematical framework.

A few practical comments are in order.  In sequential applications, numerical conditioning is very important.  Since updates of $\V S$ involve inversion and accumulation, round-off errors can degrade positive definiteness.  Cholesky-based or square-root parameterizations maintain stability.  
Furthermore, adaptive placement of inducing inputs can improve accuracy in nonstationary domains.  
The initialization of $\Vx_u$ also affects convergence: 
$k$-means clustering or Latin hypercube sampling over the input domain provide effective starting points.

\section{Other Approaches to Sequential Estimation} 

While the previous sections focused on sparse, variational, and Markovian formulations, several alternative approaches to sequential estimation have been developed in the GP literature and in related signal‐processing frameworks. These methods attempt to reconcile GP expressiveness with the computational and memory constraints of real-time applications. They include recursive kernel methods, variational state-space formulations, online expectation
propagation, and combinations of GPs with neural or particle-based estimators.  
Although conceptually diverse, they share a common goal: to maintain a compact belief state that
summarizes past information while adapting efficiently to new data.

\subsection{Recursive Kernel and Online Regression Methods}

Early work on sequential GP inference originated in the signal‐processing community through
\emph{kernel adaptive filters} \citep{engel2004kernel}. 
The kernel recursive least-squares (KRLS) algorithm and its variants \citep{van2012kernel} replace
explicit GP conditioning with recursive updates of kernel coefficients. Each new observation introduced a basis function,
and sparsification criteria determined whether it should be retained in the ``dictionary.''  
The approximate posterior mean took the form of a weighted sum of these dictionary elements, with
the corresponding covariance tracked implicitly through the weight update rule.  
Although KRLS does not maintain a full probabilistic representation, its updates are algebraically
equivalent to those of a GP with a dynamically pruned set of inducing points.  
Parallel developments in the signal-processing community established a probabilistic foundation for these methods \citep{perez2013gaussian}, leading to the \emph{kernel Kalman filter}
and \emph{GP adaptive filters} that explicitly propagated the mean and covariance
of the function values .  
These methods provided the first bridge between adaptive filtering and GP regression, and they
remain competitive in real-time nonlinear tracking and denoising tasks.

A practical distinction between kernel adaptive filters and modern sparse GPs lies in their treatment
of uncertainty.  
Kernel filters rely on deterministic or heuristic forgetting factors, whereas
variational or state-space GPs retain explicit probabilistic uncertainty through Gaussian posteriors.  
Nevertheless, the recursive kernel viewpoint remains influential, particularly in systems that require
rapid adaptation with limited computation, such as streaming sensor networks or embedded platforms.

\subsection{Variational State–Space and Hybrid Models}

A second class of methods merges GP priors with explicit state-space formulations where the transition
or observation functions are learned via neural networks.  
In these hybrid models, latent dynamics evolve according to a GP prior, but inference proceeds through
variational filtering rather than exact Kalman updates.  
The \emph{variational Kalman GP} and related frameworks \citep{hamelijnck2021spatio, dowling2023linear}
maintain a structured Gaussian posterior over both latent states and inducing variables.  
Sequential updates of the variational parameters proceed by minimizing the local Kullback–Leibler
divergence between the approximate and true filtering distributions.  
This approach inherits the expressive power of non-Markovian kernels while preserving linear-time
updates in the temporal dimension.

A practical advantage of these hybrid models is their ability to incorporate non-Gaussian likelihoods,
latent regime switches, and heteroscedastic noise without resorting to particle approximations.
They have become standard tools in modern spatiotemporal modeling, particularly in applications
involving neural activity \citep{dowling2023linear} and environmental monitoring \citep{hamelijnck2021spatio}, among others.
The combination of GP priors with amortized variational inference, implemented through recurrent
neural networks, further extends their reach to high-dimensional or irregularly sampled data streams \citep{fortuin2020gp}.

\subsection{Expectation Propagation and Message–Passing Frameworks}

Another family of sequential GP algorithms derives from expectation propagation (EP).
In the EP formulation, each data point contributes a site approximation that updates the global
posterior through moment matching.  
Sequential EP updates can be written in recursive form,
\begin{align}
q_t(f) \propto q_{t-1}(f)\,\frac{\tilde{p}_t(f_t)}{Z_t},
\end{align}
where $\tilde{p}_t(f_t)$ denotes the local Gaussian site and $Z_t$ a normalization factor.
When expressed in natural parameters, this recursion is identical to an information-form Kalman
filter with adaptive site precision.  
The \emph{state-space expectation propagation} (SSEP) method \citep{wilkinson2020state} generalizes
this idea to temporal GPs and achieves linear-time exactness for Markovian kernels and accurate
approximations for more general covariances.  
Because each update corresponds to a local moment-matching step, SSEP preserves the calibration of
posterior variances, a property often lost in purely variational schemes.

Message-passing perspectives also provide a unifying view of GP inference. From this perspective, the predictive mean and covariance evolve through forward and backward messages
that obey the same algebra as Kalman filtering but operate in function space.  
This view clarifies relationships among variational methods and EP,  and it supports extensions to hierarchical or deep GP architectures.

\subsection{Monte Carlo and Particle Approaches}

In strongly nonlinear or non-Gaussian settings, sequential estimation may rely on Monte Carlo
techniques. GP state-space models --- state-space models where the transition and/or observation models are modeled by a GP --- are a particularly challenging setting, necessitating Monte Carlo methods.

In this setting, 
\emph{GP particle filters} approximate the joint distribution over latent states and
function values by a set of weighted particles \citep{ko2009gp}.  
Each particle carries its own GP belief, which evolves through resampling and weight updates based
on predictive likelihoods.  
Although computationally expensive, these methods enable fully nonlinear and non-Gaussian inference
with GP priors and have been applied to robotics and control problems where model uncertainty must
be tracked explicitly \citep{ko2009gp}. Recent work has further incorporated alternative filtering methods into the GP SSM problem. Namely, \citet{lin2024ensemble} use the ensemble Kalman filter for efficient variational inference in GP SSM models.

\section{Comparisons to Deep Learning Models} 

GP models and deep learning represent two complementary paradigms for modeling
sequential data. Both attempt to model dependencies across time, uncertainty in prediction, and generalization to unseen conditions. Yet their philosophical foundations differ. GPs permit two complementary views. In the function–space view, the kernel defines a prior on functions, and Bayes' rule yields a posterior over functions given observed data. In the parameter–space view, finite–basis approximations such as RFF or HSGP represent the GP through a fixed set of coefficients that evolve sequentially with constant memory. Related low–rank functional approximations, such as variational inducing–point GPs, maintain the same efficiency by summarizing past information in a finite set of inducing variables. On the other hand,  deep learning constructs deterministic parametric mappings optimized for predictive accuracy. This section contrasts the sequential GP frameworks developed earlier with several major classes of deep learning methods, including recurrent and convolutional architectures, diffusion processes, and deep kernel models. We emphasize similarities in structure, differences in inductive bias, and recent unifying viewpoints that link probabilistic and neural approaches.

\subsection{Conceptual Comparison: Function–Space versus Parameter–Space Inference}

Exact GPs define inference directly in function space, where the kernel specifies a prior over
functions, and conditioning on observations produces a posterior distribution over all admissible
functions. We discussed sequentially tractable approximations such as RFF, HSGP, and variational inducing-point models, which retain the same probabilistic interpretation but express the posterior through a finite set of parameters or inducing variables. Thus, these models allow for constant-memory recursive updates with calibrated uncertainty.

Deep neural networks (DNNs), by contrast, define explicit parametric maps $f_\theta(x_t)$ with parameters $\theta$ updated in parameter space. Learning in this setting typically involves stochastic gradient descent, which yields a single point estimate of $\theta$ rather than a full distribution. Uncertainty estimates are therefore approximate and are usually obtained through ensembles, dropout sampling, or stochastic variational layers.

The contrast between fully functional inference and its finite-dimensional approximations determines how models balance data efficiency and robustness.  
Exact function-space inference relies on strong priors and remains sample-efficient, whereas parameter-space formulations such as RFF, HSGP, or variational inducing-point GPs trade some prior fidelity for scalability and numerical stability. Furthermore, GPs excel in low–data regimes, where priors constrain functional variability. By contrast, DNNs tend to require larger data volumes to learn similar inductive biases but achieve greater flexibility in high–dimensional domains. Sequential GP algorithms occupy a middle ground between analytical tractability and dynamic adaptability. They maintain explicit uncertainty through recursive
updates while scaling linearly in time.

\subsection{Diffusion Processes and the Markovian View}

A particularly interesting point of comparison arises with diffusion–based models, which have become a dominant class in modern deep learning \citep{yang2023diffusion}. Diffusion models define a forward stochastic process that gradually corrupts data with Gaussian noise and a reverse process that learns to denoise and recover the original distribution. Both directions form Markov processes in continuous time and thus bear formal similarity to the Markovian GP models discussed in Section~V. Each defines an SDE over latent trajectories, but their objectives differ: diffusion models learn (intractable) reverse dynamics by optimizing an ELBO term on the data likelihood, whereas Markovian GPs use known forward and reverse dynamics of a simpler, linear diffusion model to perform exact Bayesian inference. 

Recent work has begun to connect these two frameworks explicitly. \citet{verma2024variational} build upon the work of \citet{archambeau2007gaussian} by examining the relationship between general diffusion models (now commonplace in ML) and the LTI SDEs (described by Markovian GPs) through a variational framework. Efficient variational algorithms are developed, and they compare favorably to other variational algorithms on SDEs (such as neural SDEs). The comparison between diffusion processes and Markovian GPs is particularly interesting in that it contrasts tractable inference under a defined stochastic-process prior with more data-driven, generative deep learning models with similar mathematical foundations.

\subsection{Sequential Deep Learning Architectures}

Sequential data modeling in deep learning primarily relies on recurrent neural networks (RNNs),
long short–term memory networks (LSTMs), temporal convolutional networks (TCNs), and, more recently, transformers. These architectures propagate information through hidden states that are updated deterministically
by learned parameters. The hidden state of an RNN plays a role analogous to the latent state in a GP
state–space model, but its update follows a nonlinear map rather than a probabilistic transition.  
This distinction affects both interpretability and uncertainty quantification.

For instance, the basis expansion GP defines a linear Gaussian SSM,
\begin{align}
\V\theta_{t+1} = \V\theta_t + \varepsilon_t, \qquad y_t = h_t \V\theta_t + \eta_t,
\end{align}
where $\varepsilon_t$ and $\eta_t$ are Gaussian noise terms. The resulting filtering solution are nonlinear updates of $h_t \theta_t$ and $y_t$. An RNN replaces these stochastic transitions with a nonlinear function $f_{t+1} = \phi(W f_t + U y_t)$, where $\phi$ is a learned function and $f_t$ are a set of state variables.
While this deterministic formulation achieves strong empirical performance, it lacks the uncertainty calibration inherent in GP updates. Hybrid designs, such as variational RNNs and deep state–space models,
reintroduce latent stochasticity and thus approximate the recursive inference mechanisms of sequential GPs.

Transformers extend this idea through self–attention, which models all pairwise dependencies between time steps \citep{vaswani2017attention}. From a GP perspective, self–attention can be interpreted as defining a learned, data–dependent kernel.  
Indeed, several recent analyses interpret attention weights as low–rank approximations to a GP covariance matrix  (e.g., \citet{bui2025revisiting}). Sequential GPs with adaptive kernels, therefore, share structural parallels with transformer models but maintain an explicit probabilistic interpretation.

\subsection{Deep Kernels and Hierarchical Gaussian Processes}

Another natural point of contact between deep learning and GPs lies in kernel design.
Traditional GPs rely on hand–crafted kernels such as the squared–exponential or
Mat\'ern family, which impose fixed notions of similarity in input space.
Deep kernel learning (DKL) \citep{wilson2016deep} replaces this assumption by
parameterizing the kernel through a neural feature map,
\begin{align}
\kappa(x,x') = \kappa_0\!\left(\psi_\theta(x),\psi_\theta(x')\right),
\end{align}
where $\psi_\theta$ is a deep network and $\kappa_0$ is a base kernel such as the RBF. This composition allows the GP to operate in a learned latent representation that captures the hierarchical structure in the data while retaining Bayesian inference in the output space. 
Sequential DKL variants integrate temporal recurrence into the embedding process, exploiting the sequential nature of data in the embedding prior \citep{al2017learning}.

An alternative approach are deep Gaussian processes (DGPs) \citep{damianou2013deep}, which stack multiple GPs hierarchically, with each layter defining a random function that feeds into the next. The resulting process represents a nonparametric analog of a DNN, with explicit uncertainty propagation across layers. Although exact inference in DGPs is intractable, variational approximations yield tractable sequential updates analogous to those described in Section~VI.   From a signal–processing perspective, these models resemble cascaded dynamical systems whose intermediate outputs remain probabilistically defined. Versions of the DGP based on the Markovian GP are proposed in \citet{zhao2021deep}.

\subsection{Nonstationary and Adaptive Representations}

The intersection of deep learning and GP modeling has also led to new perspectives on nonstationarity.  
\citet{titsias2024kalman} provide an online neural network learning solution based on a method akin to a combination of DKL and the Kalman-filter-based form of linear regression. This is equivalent to taking neural network-based embeddings and a linear kernel in the basis expansion GPs previously outlined.
This approach connects neural adaptive filtering to filtering-based Bayesian models. Using the same tricks as the non-stationary GP regression discussed earlier, the model adapts dynamically to nonstationary environments while maintaining
analytic tractability in its posterior updates.

These developments suggest a continuum between purely kernel–based and purely
neural approaches.  
At one end, classical GPs provide closed–form sequential inference with strong
uncertainty calibration but limited representational flexibility.  
At the other end, deep networks provide extreme flexibility but weak uncertainty control.
Hybrid methods, including DKL, DGP, and neural–SDE formulations, offer a balance
by embedding learnable representations within probabilistic filters.

\subsection{Outlook: Complementarity and Integration}

From the viewpoint of SP, sequential GPs and deep learning differ
not as competing paradigms but as complementary tools.
GPs serve as interpretable, uncertainty–aware priors that can guide learning and
quantify confidence, whereas deep networks supply expressive, data–driven
parameterizations of complex nonlinear dependencies.
Recent work increasingly combines the two, with neural networks that amortize kernel computation, approximate marginal likelihoods, or perform variational updates. 
Conversely, GP perspectives have improved the calibration and sample efficiency of
deep generative and sequential models.

In practice, hybrid architectures that integrate GP priors with deep representations
already achieve state–of–the–art performance in time–series forecasting,
robotics, and spatiotemporal modeling.  
They preserve the recursive update structure central to SP while
drawing on the representational power of deep learning.  
This convergence underscores a broader trend: probabilistic and neural methods are
not distinct domains but parts of a unified framework for sequential inference under
uncertainty.

\section{Applications of Gaussian Processes to Sequential Signal Processing} 

The methods discussed in the preceding sections—basis expansions, Markovian representations, and
variational approximations—enable GP models to operate efficiently in sequential
environments.  In this section, we explore how these formulations translate into practical SP
applications.  We organize the discussion around several representative domains that highlight different
aspects of GP inference: sequential regression and forecasting, distributed and decentralized inference,
robust learning and anomaly detection, sequential Bayesian optimization, and adaptive sensing.  Across
these areas, GPs serve as a unifying probabilistic framework for uncertainty–aware estimation and
decision making.

\subsection{Sequential Regression and Forecasting}

Sequential regression remains one of the canonical applications of GPs in SP. In this setting, the objective is to infer a latent, possibly time–varying signal from an incoming data stream while maintaining calibrated uncertainty estimates. We saw that when combined with RFF or Hilbert–space approximations, GP regression reduces to a linear–Gaussian model that allows  recursive updates through the Kalman filter.  This formulation allows real–time prediction, filtering, and smoothing with scalable memory usage and linear time complexity in the number of observations.

Recent works demonstrate the flexibility of this approach.  
\citet{lu2022incremental} proposed incremental ensemble GPs that adapt model weights online through Bayesian model averaging.  
\citet{waxman2024doebe} extended this concept to more general basis expansions.
Other approaches focus on variational learning; \citet{chang2023memory} build upon the dual GP of \citet{adam2021dual} to include memory of past points and derive efficient sequential updates.

\subsection{Distributed and Decentralized Online Inference}

In multi–agent and networked systems, sequential GPs provide a principled mechanism for decentralized
learning.  Each node maintains a local GP model that encodes partial information about the global
process, while message passing or consensus updates ensure global coherence.  
Because the GP posterior is Gaussian, local summaries can be transmitted as information quantities --- these are equivalent to transmitting means and covariances, but decompose additively with more data --- which allows for communication-efficient fusion of information.

\citet{llorente2025decentralized} recently introduced decentralized online ensembles of
GPs, which employ the information form of the Kalman filter to propagate local sufficient
statistics.  
The framework extends classical results from distributed SP to probabilistic
nonparametric settings and allows agents to maintain uncertainty–aware beliefs without centralized
coordination.  
Related work on robust decentralized GPs~\citep{llorente2025robust} incorporates outlier resistance and
adaptive weighting and draws from robust filtering theory~\citep{chang2017unified}.  
Such approaches illustrate how sequential GP inference can support scalable, fault–tolerant estimation in
sensor networks and multi–robot systems.

\subsection{Robust Sequential Learning and Anomaly Detection}

Sequential GP models are also effective for detecting anomalies or regime changes in streaming data.  
Because GPs maintain predictive variances that reflect uncertainty about future observations, deviations
from expected predictive likelihoods naturally indicate potential outliers or distributional shifts.
Markovian and basis–expansion GPs both accommodate such analysis by tracking evolving latent states
and adapting their uncertainty in real time.

\citet{bock2022online} used state–space GPs to perform online anomaly detection and robust inference in time series, leveraging predictive residuals as test statistics for unexpected events.  
\citet{waxman2024lintel} proposed a GP–based streaming algorithm for regime
switching and outlier prediction, which use Markovian GPs for efficient updates.
More recently, \citet{laplante2025robust} introduced robust spatiotemporal GPs based on
generalized Bayesian inference, which avoid the influence of heavy–tailed noise and unmodeled
dynamics. Laplante et al. explicitly note that the sequential formulation helps resolve open problems in robust GP estimation, by allowing more informed weighting of new observations.
These developments connect GP–based filtering to long–standing interests in the SP
community, namely resilience, robustness, and principled uncertainty quantification.

\subsection{Sequential Bayesian Optimization}

GPs are widely used in Bayesian optimization due to their ability to quantify uncertainty and guide
exploration.  In sequential settings, these methods adapt the acquisition function as new data arrive, often
under real–time or resource–constrained conditions.  
Sequential GP optimization has proven valuable in adaptive control, communication systems, and
autonomous experimentation.

\citet{lu2023surrogate} proposed surrogate–model ensembles that update their posterior in real time through RFF–based approximations and achieve scalable optimization beyond a single GP.
\citet{maddox2021conditioning} introduced conditioning techniques that stabilize GP
posteriors for streaming optimization tasks while maintaining uncertainty calibration.  
By combining approximate inference with efficient online acquisition strategies, these methods preserve
the statistical rigor of Bayesian optimization while remaining computationally feasible for embedded or
interactive systems.

\subsection{Adaptive and Active Sensing}

Sequential GP inference naturally lends itself to adaptive sensing, where the goal is to select
measurements that maximize information gain or minimize uncertainty about an evolving process.
In this context, the GP posterior defines an acquisition criterion that guides sensor placement or
measurement timing.  
Because GP updates are recursive, sensor trajectories or sampling schedules can be optimized in real
time.

\citet{ziatdinov2022bayesian} demonstrated this idea for atomic–scale imaging, where
Bayesian GP frameworks adaptively guide scanning probe measurements to capture dynamic physical
phenomena.  
In spatiotemporal SP, similar strategies enable mobile sensor networks to track dynamic
fields while minimizing energy consumption and communication costs.  The probabilistic nature of GPs ensures that sensing decisions are informed by both estimated signal structure and model uncertainty, and that they align naturally with long-standing principles of active and adaptive filtering.

\section{Conclusion}

GPs offer a principled probabilistic framework that unites many classical ideas of
SP under a single mathematical formalism.  Throughout this article, we have presented
how sequential and online inference in GPs connects directly to established concepts such as recursive
least squares, Kalman filtering, and SDEs.  By introducing finite basis expansions, Markovian representations, and variational approximations, we have shown that the apparent computational limitations of GPs can be overcome while preserving their ability to quantify uncertainty.
These formulations restore the recursive structure fundamental to SP and make GPs
suitable for real-time and large-scale applications.

The equivalence between linear-Gaussian state–space models and GP priors has clarified the role of
stochastic dynamics in kernel design.  Spectral and Hilbert-space approximations translate nonparametric
GP priors into compact linear models that evolve sequentially through Kalman updates.  Markovian GPs
extend this correspondence to continuous time and yield exact inference in linear time for a wide class
of stationary kernels.  Sparse and variational formulations summarize the information contained in past observations through inducing variables or basis coefficients, which ensures constant memory and linear update complexity.  Together, these developments establish a bridge between Bayesian nonparametrics and
classical estimation theory.

The application domains discussed illustrate how sequential GPs have matured from theoretical constructs
into practical tools for modern SP.  In regression and forecasting, they provide uncertainty-aware
alternatives to deep recurrent models.  In distributed inference, they enable probabilistic cooperation across
agents through consensus updates in information form.  In robust learning and anomaly detection, they
produce calibrated likelihoods that expose distributional shifts or sensor faults.  In sequential Bayesian
optimization and adaptive sensing, they guide exploration and data acquisition under resource constraints.
Across these tasks, the GP framework provides both predictive performance and interpretability, which are two qualities seldom achieved simultaneously in contemporary ML.

Despite their progress, important research challenges remain.  Current sequential GP algorithms still
depend on stationary kernels or a limited number of basis functions, which restricts their expressiveness
in high-dimensional and strongly nonstationary systems.  More flexible kernel parameterizations, possibly
driven by neural feature maps, promise to extend their applicability while maintaining the probabilistic
interpretation that distinguishes GPs from purely data-driven models.  Another open direction concerns
scalability in multi-agent networks, where thousands of sensors or robots must coordinate their inference
under communication constraints.  Consensus-based GP filtering and distributed variational inference may
offer a path forward.  Robustness also remains an active frontier, particularly under heavy-tailed or
adversarial noise; generalized Bayesian inference and $\alpha$-divergence formulations suggest principled
solutions that retain analytic structure.

Several emerging areas invite further exploration.  In physical modeling, hybrid GP–physics approaches
link latent stochastic dynamics with partial differential equations and could redefine data assimilation in
geoscience, fluid mechanics, and biomedicine.  In signal intelligence and communications, GPs may provide
a foundation for probabilistic waveform synthesis and adaptive spectrum management, where uncertainty
quantification directly influences control decisions.  In neuroscience, GP state–space models already
enable interpretable decoding of neural activity and could evolve into real-time brain–computer interfaces
through scalable sequential inference.  In robotics, Markovian GPs integrated with diffusion or neural-SDE
representations may offer a unified framework for perception, prediction, and control under uncertainty.

From a broader perspective, GPs exemplify the convergence of statistical learning and SP.
They formalize estimation as inference in function space while maintaining the recursive and modular
character that defines signal-processing methodology.  As computing architectures continue to favor
streaming and decentralized operation, the relevance of sequential GP formulations will only increase.
Future progress will likely emerge through cross-fertilization with deep generative modeling, reinforcement
learning, and information-theoretic control, where uncertainty must be represented, propagated, and
acted upon in real time.

In summary, GPs have evolved into a versatile paradigm that extends the classical theory
of signals and systems to the probabilistic domain.  Their sequential formulations inherit the rigor of Bayesian inference and the efficiency of recursive estimation, which provides a coherent path from theory to implementation. Continued integration of these models with advances in optimization, numerical linear
algebra, and neural representation learning promises to expand their role as central tools for signal
processing in the coming decade.

\section*{Acknowledgment}
The authors would like to acknowledge the support of the National Science Foundation under Award 2212506.

%
\bibliography{main}
\bibliographystyle{imsart-nameyear}

\vfill

\end{document}